\documentclass[fleqn,usenatbib]{mnras}
\usepackage{graphicx}
\usepackage{natbib}
\usepackage{Times}
\usepackage{enumerate}
\usepackage{amsmath, amssymb}
\newcommand\lsim{~\lower.5ex\hbox{$\buildrel < \over \sim$}~}
\newcommand\gsim{~\lower.5ex\hbox{$\buildrel > \over \sim$}~}
\usepackage[T1]{fontenc}

\DeclareRobustCommand{\VAN}[3]{#2}
\let\VANthebibliography\thebibliography
\def\thebibliography{\DeclareRobustCommand{\VAN}[3]{##3}\VANthebibliography}

\usepackage{amsmath}	
\usepackage{amssymb}	
\usepackage{color}
\usepackage[dvipsnames]{xcolor}
\usepackage{comment}
\usepackage{subfig}

\title[Density waves excited by MRI in
discs]{The generation of spiral density waves by MRI in accretion discs}

\author[G.R. Mamatsashvili et al.] {G.~Mamatsashvili$^{1,2}$\thanks{E-mail:
g.mamatsashvili@hzdr.de},  D. Kobaidze$^{1,2,3}$, W. Mouhali
$^{4}$, T. Lehner$^{5}$ and L. Burdiladze$^{1,2,3}$\\
$^{1}$Helmholtz-Zentrum Dresden-Rossendorf, Bautzner Landstr. 400, D-01328 Dresden, Germany\\
$^{2}$Abastumani Astrophysical Observatory, Abastumani 0301, Georgia\\
$^{3}$Faculty of Exact and Natural Sciences, Tbilisi State University, Tbilisi 0179, Georgia\\
$^{4}$LyRIDS, ECE-Paris Engineering School, 10 Rue Sextius-Michel, 75015 Paris, France\\
$^{5}$LUTH, UMR 8102 CNRS, Observatoire de Paris-Meudon, 5 place de Janssen, F-92195 Meudon, France}
\begin{document}

\date{Accepted ?. Received  ?; in
original form  ?}

\pagerange{1--14} \pubyear{2026}

\maketitle

\begin{abstract}
We investigate the linear dynamics of non-axisymmetric perturbations in Keplerian discs subject to a weak uniform vertical magnetic field in the shearing box approximation. Perturbations are decomposed into shearing waves and evolved by numerically integrating the linearized ideal MHD equations.  The disc flow supports three basic perturbation modes: two incompressible modes — magnetic mode that undergoes magnetorotational instability (MRI) and inertia–magnetic waves — and compressible spiral density waves. The magnetic mode and inertia-magnetic waves have a low frequency of the order of Alfv\'en and orbital frequencies, respectively, while density waves have high-frequency. We introduce mode eigenfunctions and governing modal equations to analyze the dynamics of individual modes. For non-axisymmetric modes, the modal equations are coupled due to the shear of the Keplerian rotation of the disc, giving rise to a new shear-induced linear mode coupling process, which is rooted in the non-self-adjoint nature of shear flows. We focus on the generation of density waves by the dominant MRI-unstable magnetic mode. We show that initially imposed magnetic mode undergoes MRI growth and abruptly excites density waves when its radial wavenumber crosses zero. The density wave-MRI coupling is most efficient when the azimuthal and vertical wavelengths of perturbations are comparable to the disc scale height. Since density waves are compressible, whereas MRI is incompressible, this wave excitation process can also be regarded as a linear mechanism generating compressible motions via MRI-driven incompressible ones. Its implications for compressible nonzero net vertical field MRI-turbulence are also discussed.
\end{abstract}


\section{Introduction}

Understanding angular momentum and mass transport in accretion discs is one of the central problems in astrophysics, which has been substantially advanced over the past decade. Different nonmagnetic thermal processes, such as subcritical barocilic instability \citep[e.g.,][]{Klahr_bodenheimer2003, Lesur_Papaloizou2010, Lyra_Klahr2011, Klahr_etal2026} and related convective overstability \citep[e.g.,][]{Klahr_Hubbard2014, Lyra2014, Latter2016, Teed_Latter2021, Teed_Latter2025}, Rossby wave instability \citep[e.g.,][]{Lovelace1999, Meheut_etal2012, Lyra_MacLow2012,Miranda_etal2016,Ono_etal2018}, vertical shear instability \citep[e.g.,][]{Nelson_etal2013,Stoll_Kley2014,Barker_Latter2015,Manger_etal2020, Shariff_Umurhan2024, Lesur_etal2025, Ogilvie2025} as well as magnetohydrodynamic (MHD)  processes, which are first of all magnetorotational instability \cite[MRI,][]{Balbus_Hawley1991, Balbus_Hawley1992}, Hall-shear instability \citep[e.g.,][]{Kunz2008, Kunz_Lesur2013, Lesur_etal2014, Simon_etal2015, Bethune_etal2017} and disc winds \citep[e.g.,][]{Wang_etal2019, Gressel_etal2020, Lesur2021} been explored in greater detail and characterized their ability to transport angular momentum \cite[see a recent review][]{Lesur_etal2023}. Nevertheless, the MRI remains one of the most robust mechanisms for destabilizing weakly magnetized Keplerian discs and driving MHD turbulence that provides angular momentum transport at rates broadly consistent with observationally inferred accretion timescales \citep[e.g.][]{Meheut_etal2015, Salvesen_etal2016, Flock_etal2017, Ross_Latter2018, Gogichaishvili_etal2018,Begelman_Armitage2023}.

In addition to turbulence, spiral density waves (SDWs) play a significant role in driving dynamical activity and angular momentum transport in discs as well as determining their morphology \citep{Muto_etal2012,Benisty_etal2015,Perez_etal2016,Dong_etal2018,Huang_etal2019}. They arise from the gas compressibility and can be excited by various mechanisms, including self-gravity \cite[e.g.,][and references therein]{Goldreich_Tremaine1978,Durisen_etal2007,Kratter_Lodato2016,Bethune_Kley2021}, vortices \citep{Bodo_etal2005, Mamatsashvili_Chagelishvili2007, Mamatsashvili_Rice09, Heinemann_Papaloizou2009a, Paardekooper_etal2010} and rotating planets embedded in the disc \citep{Goodman_Rafikov2001,Rafikov2002, Bae_Zhu2018, Cimerman_Rafikov2021}. SDWs are capable of enhancing outward transport of angular momentum caused by other mechanisms and are even a dominant mode responsible for this transport in self-gravitating discs \citep{Kratter_Lodato2016}. Moreover, SDWs generated in the active layers in protoplanetary discs can penetrate into the poorly ionized ``dead zones'', where they can in turn drive angular momentum transport \citep{Oishi_MacLow2009}. In disc–planet interactions, a planet excites a pair of SDWs that propagate radially inward and outward in the disc. These waves exert torques on the planet, modifying its migration rate. 

The dynamics of SDWs in magnetized discs and, in particular, their interplay with MRI has also been considered. \cite{Heinemann_Papaloizou2009b} studied the excitation of SDWs in MRI-turbulence in discs with zero net magnetic flux in the shearing box approximation. In this case, the SDW generation is in fact hydrodynamic rather than MHD by nature, has a linear origin and is associated with a vertically uniform vortical mode with non-zero potential vorticity that acts as a source of SDWs, consistent with their original linear analysis \citep[][hereafter HP09a]{Heinemann_Papaloizou2009a}. Using a temporal WKB approach and matching the asymptotic solutions for the initially imposed vortical mode to that for the subsequently excited SDWs, these authors quantified the efficiency of SDW generation by the vortical mode as a function of azimuthal wavenumber. It turned out to be most efficient when the azimuthal wavenumber of SDWs is comparable to the disc scale height, while these waves, like the vortical mode, do not vary vertically. On the other hand, magnetic field perturbations appeared to play only a minor role in the wave generation process. However, this might be due to the specific configuration considered in HP09a -- a weak magnetic field with zero net flux -- which prevents the development of more dominant, large-scale, exponentially growing MRI modes.

 Large-scale regular wave structures nearly uniform along the vertical direction have been often observed in the density field of compressible MRI-driven turbulence with nonzero net flux in other simulations \cite[e.g., ][]{Fromang_Stone2009, Blaes_etal2011, Gressel_etal2012,Bai_Stone2013,Zhu_etal2013, Ross_Latter2018, Sun_Bai2021}. These wave-like density features have been attributed to SDWs produced by vortex-wave coupling mechanism of HP09a in MRI-turbulence. However, unlike in the zero net flux case, growing MRI modes in the nonzero net flux case can dominate the vortical mode, thereby influencing SDW generation and dynamics. The detailed mechanism underlying the coupling between magnetic field perturbations and SDWs is still not well understood.

 The linear coupling of vortices and SDWs in discs revealed by HP09a is in fact a special manifestation of a general linear mode coupling phenomenon in shear flows that arises from their non-self-adjoint nature \citep{Trefethen_etal1993}. The non-self-adjoitness of shear flows and its first consequence -- transient, or nonmodal growth of perturbations were well understood and described in detail by the hydrodynamic community in the 1990s \citep[][and references therein]{Schmid_Henningson2001, Schmid2007}. Later, the relevance of the nonmodal growth phenomenon in the dynamics of accretion discs, which are a special type of shear flows, and transition to turbulence therein was also demonstrated \citep{Lominadze_etal1988,Chagelishvili_etal03, Zhuravlev_Razdoburdin2014}. A second significant consequence of non-self-adjoitness of shear flows is the linear mode coupling phenomenon, which was revealed first as a coupling of vortical and sound waves in hydrodynamic shear flows \citep{Chagelishvili_etal1997} and later as a coupling between different wave branches in MHD shear flows \citep{Chagelishvili_etal96, Gogoberidze_etal2004}. It plays a central  role in the subsequent nonlinear evolution of perturbations and largely defines the characteristics of the resulting nonlinear state (turbulence).

 The shear-induced linear mode coupling phenomenon inevitably occurs in discs because of their strong Keplerian differential rotation, or shear. Indeed, several manifestations of this coupling process in discs have been analyzed in detail. First of all it is the generation of SDWs by vortices in Keplerian discs,  which was originally studied by \cite{Bodo_etal2005} both in local and global 2D non-self-gravitating discs and generalized by \cite{Mamatsashvili_Chagelishvili2007} to self-gravitating discs, where it is much more efficient. Later, as mentioned above, this process was discussed in zero net flux MRI-turbulence in magnetized discs by \cite{Heinemann_Papaloizou2009b}. Other notable manifestations of the linear mode coupling in discs include generation of inertia-gravity waves by vortices \citep{Tevzadze_etal2003,Tevzadze_etal08}, generation of SDWs by entropy and vortical modes in the presence of  baroclinic instability \citep{Tevzadze_etal10}, generation of SDWs by convection \citep{Mamatsashvili_Rice2011} and generation of inertia-gravity waves by MRI-unstable modes \citep{Mamatsashvili_etal2013}.

In this paper, we study the linear dynamics and coupling of SDW and MRI modes in compressible Keplerian discs subject to a vertical uniform background magnetic field in the shearing box approximation. This extends, on the one hand, the work of HP09a to the case of nonzero net vertical magnetic flux and, on the other hand, our previous study \cite{Mamatsashvili_etal2013}, in which we demonstrated the linear coupling between inertial–gravity waves and MRI modes in incompressible stratified discs. The zero and nonzero net flux configurations are two main ones where MRI is usually studied. The MRI dynamics is different in these two cases. In the first case, there is  a well-defined linear instability leading to the exponential growth of axisymmetric (channel) modes, which are most unstable, and the transient, or nonmodal growth of non-axisymmetric modes over a finite time \citep{Goodman_Xu1994, Pessah_Goodman2009, Latter_etal2009, Latter_etal2010, Mamatsashvili_etal2013}. On the other hand, in the second case, MRI sets in as a subcritical instability and nonlinear dynamo action must be effective to sustain the large-scale field that in turn supports MRI \citep{Herault_etal2011,Riols_etal2017, Rincon2019}. As a result, MRI is stronger in the nonzero net flux case -- it is characterized by higher growth rate and hence nonlinearly saturates into MHD turbulence with a higher transport rate -- than that in the zero-net flux case. This should clearly have an effect on the SDW generation process, since the main source responsible for wave generation can be magnetic field rather than potential vorticty, which is anyway not conserved in the 3D MHD case. By contrast, as shown by \cite{Heinemann_Papaloizou2009b}, in zero net flux MRI-turbulence, a magnetic field was not actually important in the SDW generation process, which, as mentioned above, is essentially of hydrodynamic nature stemming from potential vorticity. We show here that non-axisymmetric modes, undergoing nonmodal MRI growth, can excite high-frequncy SDWs due to shear-induced linear mode coupling process and identified a range of azimuthal and vertical wavenumbers where this SDW-MRI coupling is most efficient. 

The paper is organized as follows. In Section 2, we describe the mathematical formalism -- linearize basic ideal MHD equations, classify modes and cast those equations into a system of six first order differential equations for mode eigenfunctions. Section 3 is devoted to the dynamics of non-axisymmetric modes, demonstrating the generation of SDWs by the magnetic mode and examining its efficiency as a function of wavenumbers. Summary and discussion are given in Section 4.

\section{Physical model and main equations}

We use the shearing box approximation
\citep{Goldreich_Lynden-Bell65, Hawley_etal1995} to study the dynamics of perturbations in a magnetized Keplerian disc. In this model, gas motion is considered in a local Cartesian frame $(x,y,z)$, where $x$ is the radial, $y$ is the azimuthal and $z$ is the vertical coordinate with the corresponding unit vectors $({\bf \hat{x}}, {\bf \hat{y}}, {\bf \hat{z}})$ in these directions. This frame co-rotates with the disc's angular velocity $\Omega$  at some fiducial radius $r_0$ from the central object. In this local approach, curvature effects due to cylindrical geometry
of the disc are ignored.  For simplicity, we neglect stratification  due to the vertical gravity of the central object, because it is not important for the shear-induced mode coupling phenomenon studied here. In this unstratified shearing box, the equations of compressible ideal MHD are \citep[e.g.,][]{Hawley_etal1995}:
\begin{equation}
\frac{\partial\rho}{\partial t}+\nabla\cdot({\rho\bf u})=0,\label{eq:mass}
\end{equation}
\begin{multline}
\frac{\partial {\bf u}}{\partial t}+({\bf u}\cdot\nabla){\bf
u}+2\Omega{\bf\hat{z}}\times {\bf u}= -\frac{1}{\rho}\nabla
\left(p+\frac{{\bf B}^2}{8\pi}\right)\\+\frac{({\bf B}\cdot
\nabla){\bf B}}{4\pi\rho}+2q\Omega^2x{\bf \hat{x}} \label{eq:motion}
\end{multline}
\begin{equation}
\frac{\partial {\bf B}}{\partial t}=\nabla \times ({\bf u}\times
{\bf B}),\label{eq:induction}
\end{equation}
\begin{equation}
\nabla\cdot {\bf B}=0, \label{eq:divB}
\end{equation}
where ${\bf u}$ is the velocity in the local frame, $\rho$ and $p$ are, respectively, the gas density and pressure, and ${\bf B}$ is the magnetic field. We close this system with an isothermal equation of state for the gas, $p=c_s^2\rho$, where $c_s$ is the constant sound speed. 

In this rotating frame, the unperturbed Keplerian rotation of the disc represents an azimuthal flow ${\bf u_0}=-q\Omega x {\bf \hat{y}}$ with a constant velocity shear in the radial direction, where the shear parameter $q=1.5$ for the Keplerian rotation. The equilibrium density $\rho_0$ and the pressure $p_0$ are spatially uniform due to the absence of vertical gravity. The flow is subject to a uniform vertical background magnetic field ${\bf
B}_0=B_{0z}{\bf \hat{z}}$, which is assumed to be weak (sub-thermal)
in the sense that the corresponding Alfv\'{e}n speed,
$v_{A}=B_{0z}/(4\pi \rho_0)^{1/2}$, is much smaller than the sound
speed $c_s$, i.e., the usual plasma $\beta=2c_s^2/v_{A}^2\gg 1$.
Below we take $\beta=400$ comparable to that adopted in MRI studies in discs.

\subsection{Perturbation equations}

Consider small perturbations of density $\rho'=\rho-\rho_0$, velocity ${\bf u}'={\bf u}-{\bf u}_0$ and magnetic field ${\bf b}'={\bf B}-{\bf B}_0$ about the equilibrium state described above. Linearizing equations (\ref{eq:mass})-(\ref{eq:divB}), we obtain the following system that governs the perturbation dynamics
\begin{equation}
\frac{D\rho'}{Dt}+\rho_0\nabla\cdot{\bf u'} =0, \label{eq:per_rho}
\end{equation}
\begin{equation}
\frac{Du'_x}{Dt}=-\frac{1}{\rho_0}\frac{\partial}{\partial
x}\left(c_s^2\rho'+\frac{B_{0z}b'_z}{4\pi}\right)+2\Omega
u'_y+\frac{B_{0z}}{4\pi\rho_0}\frac{\partial b'_x}{\partial z},\label{eq:per_ux}
\end{equation}
\begin{equation}
\frac{Du'_y}{Dt}=-\frac{1}{\rho_0}\frac{\partial}{\partial
y}\left(c_s^2\rho'+\frac{B_{0z}b'_z}{4\pi}\right)+(q-2)\Omega
u'_x+\frac{B_{0z}}{4\pi\rho_0}\frac{\partial b'_y}{\partial z},\label{eq:per_uy}
\end{equation}
\begin{equation}
\frac{Du'_z}{Dt}=-\frac{1}{\rho_0}\frac{\partial}{\partial
z}\left(c_s^2\rho'+\frac{B_{0z}b'_z}{4\pi}\right)+\frac{B_{0z}}{4\pi\rho_0}\frac{\partial
b'_z}{\partial z},\label{eq:per_uz}
\end{equation}
\begin{equation}
\frac{Db'_x}{Dt}=B_{0z}\frac{\partial u'_x}{\partial z},\label{eq:per_bx}
\end{equation}
\begin{equation}
\frac{Db'_y}{Dt}=B_{0z}\frac{\partial u'_y}{\partial z}-q\Omega
b'_x, \label{eq:per_by}
\end{equation}
\begin{equation} 
\frac{\partial b'_x}{\partial x}+ \frac{\partial b'_y}{\partial y} + \frac{\partial b'_z}{\partial z}=0,\label{eq:per_divb}
\end{equation}
where
\[
\frac{D}{Dt} \equiv \frac{\partial}{\partial t}-q\Omega x
\frac{\partial}{\partial y},
\]
is the derivative along the base Keplerian shear flow. The form of equations (\ref{eq:per_rho})-(\ref{eq:per_divb}) permits a decomposition of the
perturbations into shearing plane waves with time-dependent amplitudes and phases, which are the natural basis functions  in the shearing box \citep{Goldreich_Lynden-Bell65, Hawley_etal1995}
\begin{equation}
F({\bf r},t)=\bar{F}(t){\rm exp}[{\rm i}k_x(t)x+{\rm i}k_yy+{\rm
i}k_zz], \label{eq:Fourier}
\end{equation}
where $F\equiv (\rho',{\bf u}',{\bf b}')$ denotes the perturbations and $\bar{F}\equiv(\bar{\rho}, \bar{\bf u},\bar{{\bf
b}})$ their Fourier amplitudes. The azimuthal,
$k_y$, and vertical, $k_z$, wavenumbers do not change with time, while the radial wavenumber $k_x(t)=q\Omega k_y t$ varies with time due to shear at a constant rate $q\Omega k_y$ for non-axisymmetric modes with $k_y\neq 0$. For convenience, in $k_x(t)$ we have shifted the origin
of time towards negative values, so that $k_x(0)=0$ at $t=0$. In this case, any initially leading shearing wave with $k_x(t)/k_y < 0$
at $t<0$, eventually changes (swings) to a trailing orientation with $k_x(t)/k_y > 0$ at
$t>0$ as its $k_x$ increases.

Substituting (\ref{eq:Fourier}) into equations (\ref{eq:per_rho})-(\ref{eq:per_divb}), we arrive at the
following system of first order ordinary differential equations that
govern the linear dynamics of Fourier amplitudes of perturbations
\begin{equation}
\frac{d\bar{\rho}}{dt}+{\rm
i}\rho_0\left[k_x(t)\bar{u}_x+k_y\bar{u}_y+k_z\bar{u}_z\right]=0, \label{eq:rhok}
\end{equation}
\begin{equation}
\frac{d\bar{u}_x}{dt}=-\frac{{\rm
i}k_x(t)}{\rho_0}\left(c_s^2\bar{\rho}+\frac{B_{0z}\bar{b}_z}{4\pi}\right)+2\Omega
\bar{u}_y+\frac{{\rm i}k_zB_{0z}}{4\pi\rho_0}\bar{b}_x, \label{eq:uxk}
\end{equation}
\begin{equation}
\frac{d\bar{u}_y}{dt}=-\frac{{\rm
i}k_y}{\rho_0}\left(c_s^2\bar{\rho}+\frac{B_{0z}\bar{b}_z}{4\pi}\right)+(q-2)\Omega
\bar{u}_x+\frac{{\rm i}k_zB_{0z}}{4\pi\rho_0}\bar{b}_y, \label{eq:uyk}
\end{equation}
\begin{equation}
\frac{d\bar{u}_z}{dt}=-\frac{{\rm
i}k_z}{\rho_0}c_s^2\bar{\rho}, \label{eq:uzk}
\end{equation}
\begin{equation}
\frac{d\bar{b}_x}{dt}={\rm i}k_zB_{0z}\bar{u}_x, \label{eq:bxk}
\end{equation}
\begin{equation}
\frac{d\bar{b}_y}{dt}={\rm i}k_zB_{0z}\bar{u}_y-q\Omega \bar{b}_x, \label{eq:byk}
\end{equation}
\begin{equation}
k_x(t)\bar{b}_x+k_y\bar{b}_y+k_z\bar{b}_z=0. \label{eq:divbk}
\end{equation}
Eliminating velocities from equations (\ref{eq:uzk})-(\ref{eq:byk}), we rewrite
equations (\ref{eq:rhok}-(\ref{eq:divbk}) only for the density $\bar{\rho}$ and the field components $\bar{b}_x$ and $\bar{b}_y$ (henceforth bars over the Fourier amplitudes will be omitted),
\begin{multline}
\frac{d^2\rho}{dt^2}+c_s^2k^2(t)\rho=(v_A^2k^2(t)-2q\Omega^2)\frac{k_x(t)}{k_z}b_x+v_A^2k^2(t)\frac{k_y}{k_z} b_y\\-2\Omega \frac{k_x(t)}{k_z}\frac{db_y}{dt}-2(q-1)\Omega \frac{k_y}{k_z}\frac{db_x}{dt}, \label{eq:rhokk}
\end{multline}
\begin{multline}
\frac{d^2b_x}{dt^2}=2\Omega\frac{db_y}{dt}+\left[2q\Omega^2-v_{A}^2(k_x^2(t)+k_z^2)\right]b_x\\-v_A^2k_x(t)k_yb_y+c_s^2k_x(t)k_z\rho, \label{eq:bxkk}
\end{multline}
\begin{multline}
\frac{d^2b_y}{dt^2}=-2\Omega\frac{db_x}{dt}-v_{A}^2(k_y^2+k_z^2)b_y\\-v_A^2k_x(t)k_yb_x+c_s^2k_yk_z\rho, \label{eq:bykk}
\end{multline}
where $k^2(t)=k_x^2(t)+k_y^2+k_z^2$ and without loss of generality we assume $k_y,k_z\geq 0$. Equations (\ref{eq:rhokk})-(\ref{eq:bykk}) form the basis for our subsequent analysis. They describe the dynamics and coupling of different perturbation modes existing in the considered compressible Keplerian flow  with a vertical background field in the shearing box model. We will classify these modes below using the Wentzel–Kramers–Brillouin (WKB) approach with respect to time. 

Combining equations (\ref{eq:divbk}), (\ref{eq:bxkk}) and (\ref{eq:bykk}), we can further simplify equation (\ref{eq:rhokk}) to a more compact form,
\begin{equation}
\frac{d^2\rho}{dt^2}+c_s^2k_z^2\rho=\frac{d^2b_z}{dt^2}, \label{eq:bzkk}
\end{equation}
which will be useful in the following WKB analysis.

\subsection{WKB approach and classification of modes}

The shear $q$ enters equations (\ref{eq:rhokk})-(\ref{eq:bykk}) implicitly
through the time-dependent radial wavenumber $k_x(t)=q\Omega k_y t$ only for non-axisymmetric, $k_y\neq 0$, modes and explicitly on the right hand side of equations (\ref{eq:rhokk}) and (\ref{eq:bxkk}). Therefore, we can
distinguish between two types of shear-induced effects. The implicit dependence on the shear leads
to finite-time linear phenomena -- transient, or nonmodal growth and linear mode coupling -- in the dynamics of non-axisymmetric perturbations that arise from the drift of the time-varying $k_x(t)$ along the $k_x$-axis and vanish for axisymmetric, $k_y=0$, perturbations. The explicit dependence on the shear, on the other hand, affects both axisymmetric and non-axisymmetric perturbations. It sets the frequency of epicyclic  oscillations and, importantly, is responsible for the existence of MRI in differentially rotating magnetized discs \citep{Balbus_Hawley1991,Balbus_Hawley1998}.

To classify the modes present in the given magnetized disc flow, we first consider the case of large wavenumbers $kH \gg 1$, where $H=c_s/\Omega$ is the disc scale height, and  $k_y \ll k_z$ when the WKB, or adiabatic approximation holds uniformly at all times  \citep{Balbus_Hawley1992, Johnson2007}. This means that the time-variation of $k_x$ and hence the implicit dependence on the shear in equations (\ref{eq:rhokk})-(\ref{eq:bykk}) can be neglected. We will see below that at larger $k_y$ and smaller $k_x(t)$ which ensues eventually as $k_x(t)$ decreases in absolute value and crosses the point $k_x=0$, the mode dynamics is no longer adiabatic and hence nonmodal shear-induced effects come into play. 

In this WKB regime, we look for solutions with a standard modal form $\rho, b_x, b_y \propto {\rm
exp}\left(-{\rm i}\int^t\omega(t') dt'\right)$. Substituting it into equations (\ref{eq:bxkk})-(\ref{eq:bzkk}), using equations (\ref{eq:divbk}) to eliminate $b_z$ and assuming that the time-dependent frequency, $\omega(t)$, satisfies the adiabatic approximation
$|d\omega(t)/dt|\ll\omega^2(t)$, we obtain the linear algebraic equations
\begin{equation}
\omega^2(k_xb_x+k_y b_y)+k_z(\omega^2-c_s^2k_z^2)\rho=0,\label{eq:algebraic1}
\end{equation}
\begin{multline}
\left[\omega^2+2q\Omega^2-v_{A}^2(k_x^2+k_z^2)\right]b_x\\
-(2{\rm i}\omega\Omega+v_A^2k_xk_y)b_y+c_s^2k_xk_z\rho=0,\label{eq:algebraic2}
\end{multline}
\begin{multline}
(2{\rm i}\omega\Omega-v_A^2k_xk_y)b_x \\+ \left[\omega^2-v_A^2(k_y^2+k_z^2)\right]b_y+c_s^2k_yk_z\rho=0. \label{eq:algebraic3}
\end{multline}
Equating the determinant of this linear system to zero yields the following sixth-order dispersion relation
\begin{equation}
\omega^6-a_1\omega^4+a_2\omega^2+a_3=0, \label{eq:dispersion}
\end{equation}
with the coefficients
\[
a_1=2(2-q)\Omega^2+\left(c_s^2+v_{A}^2\right)k^2+v_{A}^2k_z^2,
\]
\[
a_2=v_{A}^2(v_{A}^2+2c_s^2)k_z^2k^2+4\Omega^2c_s^2k_z^2-2q\Omega^2(c_s^2+v_{A}^2)(k_y^2+k_z^2)
\]
\[
a_3=c_s^2v_{A}^2k_z^2\left[2q\Omega^2(k_y^2+k_z^2)-v_{A}^2k_z^2k^2\right].
\]
For horizontally uniform $k_x=k_y=0$ perturbations, dispersion relation (\ref{eq:dispersion}) coincides with that of the compressible single-fluid one in \citet[][see their equation 61]{Blaes_Balbus1994}. In the incompressible limit $c_s\rightarrow \infty$, at $k_y=0$ it reduces to the well-known dispersion relation of \cite{Balbus_Hawley1991} that describes the axisymmetric MRI \cite[see also][]{Johnson2007, Mamatsashvili_etal2013}. When shear is absent $q=0$, this dispersion relation describes classical MHD waves in a rotating fluid \citep{Lehnert1954}. 
 So, we can view equation (\ref{eq:dispersion}) as a generalized WKB dispersion relation. It is a cubic polynomial for $\omega^2$ with three distinct solutions, which we denote  in ascending order by $\omega^2_m < \omega^2_{im} < \omega^2_{s}$, 
corresponding to three physically different types of perturbation modes. To clarify the nature of these modes, using the limit $kH\gg 1$, we derive approximate analytical solutions to equation (\ref{eq:dispersion}). These solutions have simpler, more physically transparent forms and describe the following modes:
\begin{enumerate}[1.]
\item
High-frequency compressive \textit{SDWs} with
\begin{equation}
\omega_s^2 \approx a_1=2(2-q)\Omega^2+\left(c_s^2+v_{A}^2\right)k^2+v_{A}^2k_z^2, \label{eq:freqSDW}
\end{equation}
the restoring force for which is provided by pressure (compressibility), but is modified by rotation and magnetic field. Since the background field is weak ($c_s\gg v_{A}$), expression (\ref{eq:freqSDW}) reduces to the well-known local dispersion relation of SDWs in non-magnetized discs, $\omega_s^2=2(2-q)\Omega^2+c_s^2k^2$ \citep{Goldreich_Tremaine1978,Balbus2003}.
\item
Low-frequency \textit{inertia-magnetic waves (IMWs)} with
\begin{equation}
\omega^2_{im}= \frac{a_2+\sqrt{a_2^2+4a_1a_3}}{2a_1}, \label{eq:freqIMW}
\end{equation}
the restoring force for which is provided by Coriolis and magnetic tension forces but is modified by compressibility. In the incompressible limit, for a weak magnetic field, $v_{A}k_z\ll \Omega$, and for $k_y=0$, these waves become usual inertial waves with $\omega^2_{im}=2(2-q)\Omega^2k_z^2/k^2$ \citep{Balbus2003, Johnson2007}, while for a strong magnetic field $v_Ak_z\gg \Omega$ they oscillate with the Alfv\'{e}n frequency, $\omega_{in}=v_Ak_z$.  Since $\omega^2_{im}>0$, IMWs are stable. At $k_z=0$ they reduce to stationary, vertically uniform, hydrodynamic 2D vortical modes. 
\item
\textit{Magnetic mode} with frequency/growth rate\footnote{In the literature, this mode is often called MRI mode, but we prefer to refer to it as a magnetic mode \citep[see also][]{Ogilvie98}, because besides MRI at smaller $k_x$, it is stable at large $|k_x|$ and reduces to Alfv\'{e}n wave.}
\begin{equation}
\omega^2_m=\frac{a_2-\sqrt{a_2^2+4a_1a_3}}{2a_1}, \label{eq:magmode}
\end{equation}
the restoring force for which is provided by a magnetic tension force but is modified by compressibility and rotation. That is, this mode arises from the presence of the background field, because, as follows from equation (\ref{eq:magmode}), $\omega^2_m \propto v_A^2$ and therefore disappears when $v_{A}\rightarrow 0$. The magnetic mode plays an important dynamical role because it is the one that exhibits MRI. Indeed, it follows from equation (\ref{eq:magmode}) that the magnetic mode is unstable, i.e., $\omega_m^2<0$, if the coefficient $a_3>0$, which implies
\begin{equation}
v_{A}^2k_z^2 < 2q\Omega^2,~~~~~~k_x^2<(2q\Omega^2-v_{A}^2k_z^2)\frac{k_y^2+k_z^2}{v_{A}^2k_z^2}. \label{eq:condMRI}
\end{equation}
On the other hand, at large $|k_x|$ and/or smaller shear $q$, when $a_3<0$, this mode is stable, $\omega_m^2\geq 0$, reducing to Alfv\'en wave. 
\end{enumerate}

\begin{figure}
\includegraphics[scale=0.5]{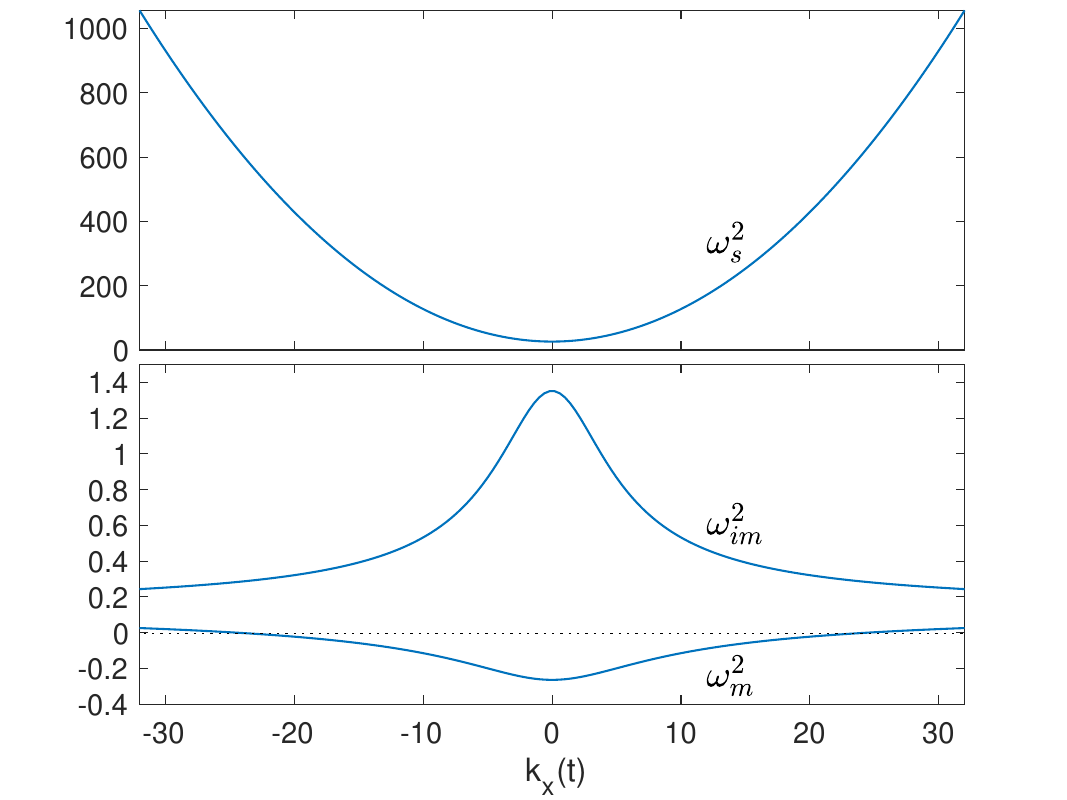}
\caption{Squared frequencies $\omega^2_s$ of SDWs,  $\omega^2_{im}$ of IMWs and $\omega^2_m$ of the magnetic mode as a function of $k_x(t)$ at $k_y=1, k_z=5$. $\omega^2_m$ is negative in the range of $k_x$ given by equation (\ref{eq:condMRI}), indicating MRI of the magnetic mode in this range with the growth rate $\gamma=\sqrt{-\omega_m^2}$, which reaches its maximum, $\gamma_m=0.51$, at $k_x=0$.}\label{fig:omega2}
\end{figure}

Figure \ref{fig:omega2} shows the squared frequencies of SDW, IMW and magnetic mode as a function of $k_x$ (which itself changes with time) for a given $k_y$ and $k_z$. $\omega^2_{im}$ and $\omega^2_{s}$ are always positive, i.e., these modes are stable, whereas $\omega^2_m<0$ in a certain range of $k_x$ given by conditions (\ref{eq:condMRI}), which indicates MRI of the magnetic mode with the growth rate $\gamma=\sqrt{-\omega_m^2}$, reaching its maximum at $k_x=0$. At large $|k_x|\gg k_y, k_z$ the effects of rotation and shear become small in the dynamics of IMWs and the magnetic mode and their oscillation frequencies approach the Alfv\'{e}nic frequency, $\omega_m \approx \omega_{im}=v_{A}k_z$. 

In this figure and everywhere below, we non-dimensionalize time by $\Omega^{-1}$, frequency by $\Omega$, length by $H$, wavenumber by $H^{-1}$, velocity by $c_s$, density by $\rho_0$ and magnetic field by $B_{0z}$. The non-dimensional Alfv\'en speed is then $v_A=\sqrt{2/\beta}$.

Thus, perturbation modes in a compressible, unstratified, weakly magnetized disc can be classified into three basic types -- compressible SDWs and two incompressible IMW and the magnetic modes. Below we focus on the shear-induced linear coupling between the MRI-unstable, and hence the strongest, magnetic mode with SDWs. To this end, following \cite{Gogoberidze_etal2004, Tevzadze_etal10,Mamatsashvili_Rice2011,
Mamatsashvili_etal2013}, we introduce modal eigenfunctions, which can be convenient when studying mode dynamics in the presence of shear.

Switching to new variables, 
\[
h_1=\rho-b_z,~~h_2=\frac{d}{dt}(\rho-b_z),~~
h_3=b_x,
\]
\[
h_4=\frac{db_x}{dt},~~~
h_5=b_y,~~~h_6=\frac{db_y}{dt},
\]
substituting them into equations (\ref{eq:bxkk})-(\ref{eq:bzkk}) and using the WKB-type modal time-dependence given above, we obtain 
\begin{equation}
\frac{d{\bf h}}{dt}=-i\omega{\bf h}={\bf A}\cdot {\bf h},\label{eq:matrix_eqs}
\end{equation}
which is essentially the same as the linear system of algebraic equations (\ref{eq:algebraic1})-(\ref{eq:algebraic3}), but rewritten in a matrix form with ${\bf h}=[h_1, h_2, h_3, h_4, h_5, h_6]^T$ being the state column vector
and the evolution  matrix ${\bf A}$ is given in Appendix A. The eigenvalues of the linear system (\ref{eq:matrix_eqs}) are found from
\[
{\rm Det}({\bf A}+{\rm i}\omega{\bf I})=0,
\]
which yields the same dispersion relation (\ref{eq:dispersion}) for $\omega$ and hence its roots, multiplied by $-{\rm i}$, $\pm {\rm i}\omega_m$,$\pm
{\rm i}\omega_{im}$ and $\pm {\rm i}\omega_s$, represent the eigenvalues of the matrix ${\bf A}$. Since these eigenvalues are generally distinct and the corresponding eigenvectors
are linearly independent, we can perform the eigendecomposition of ${\bf A}$  \citep{Golub_VanLoan96},
\begin{equation}
{\bf A}={\bf C}{\bf \Lambda}{\bf C^{-1}}, \label{eq:eigendecomp}
\end{equation}
where ${\bf \Lambda}$ is the diagonal matrix whose elements are the corresponding eigenvalues 
\[
{\bf \Lambda}={\rm diag}({\rm i}\omega_m, -{\rm i}\omega_m, {\rm i}\omega_{im}, -{\rm i}\omega_{im}, {\rm i}\omega_s,-{\rm i}\omega_s)
\]
and the columns of the matrix ${\bf C}$ are the eigenvectors of ${\bf A}$ ordered such that the first column corresponds to the eigenvalues ${\rm i}\omega_m$, second to $-{\rm i}\omega_m$, third to ${\rm
i}\omega_{im}$, fourth to $-{\rm i}\omega_{im}$, fifth to
${\rm i}\omega_s$ and sixth to $-{\rm i}\omega_s$, which are all explicitly given in Appendix B.

\begin{figure}
\includegraphics[scale=0.41]{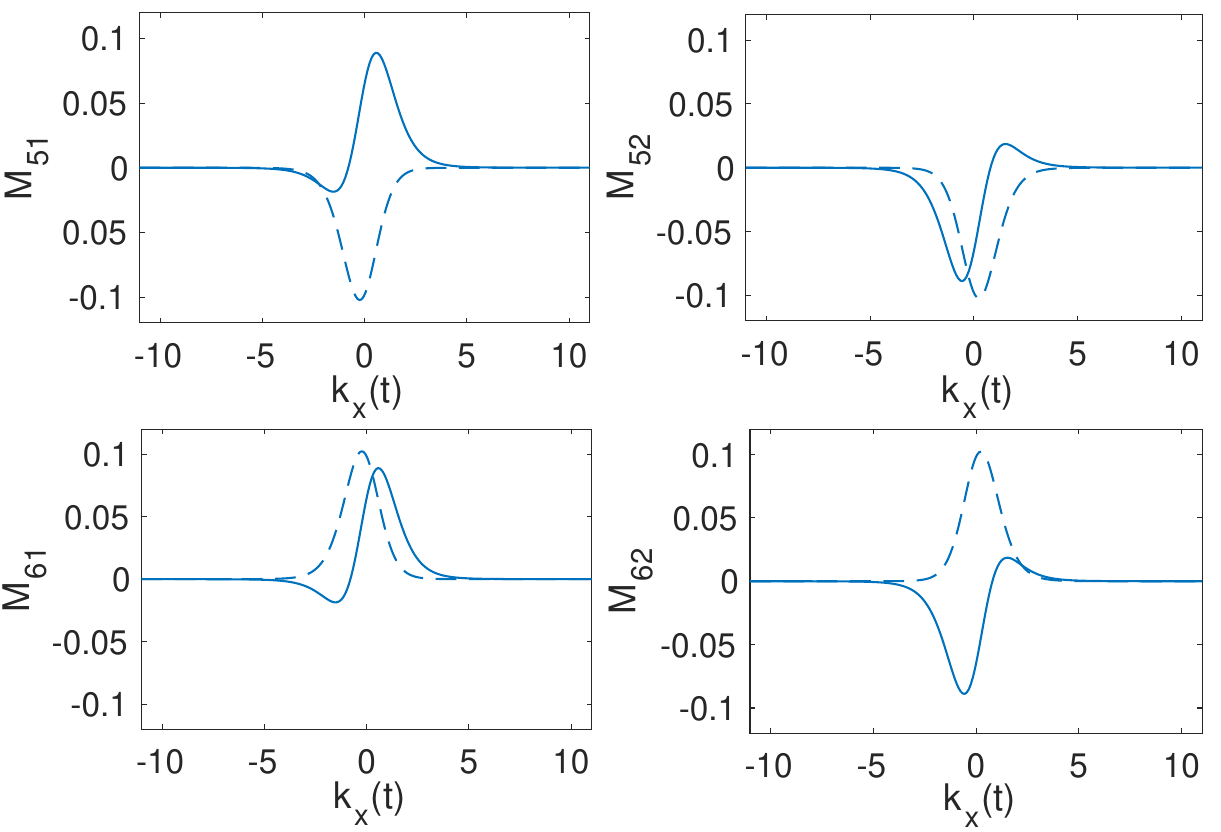} 
\caption{
The non-diagonal elements $M_{51},M_{52}$ and $M_{61},M_{62}$ of the coupling matrix ${\bf M}$, which are responsible for the coupling of the magnetic mode and SDWs, as a function of $k_x(t)$ at $k_y=2,~k_z=1$ (the solid lines show real and the dashed lines imaginary parts). These coupling coefficients vanish at large $|k_x(t)|\gg k_y$, in the WKB regime, where the dynamics is adiabatic and the modes are decoupled, but are appreciable in the non-adiabatic regime at $|k_x(t)|\lsim k_y$, resulting in the mode interactions.}\label{fig:coupling_coeff}
\end{figure}

Substituting (\ref{eq:eigendecomp}) into equation (\ref{eq:matrix_eqs}), we obtain a canonical form of this system that consists of a set of first order differential equations governing each of these three modes, 
\begin{equation}
\frac{d{\boldsymbol \psi}}{dt}={\bf \Lambda}\cdot {\boldsymbol
\psi}, \label{eq:modal_eqs}
\end{equation}
where ${\boldsymbol \psi}\equiv[\psi^{(+)}_m, \psi^{(-)}_m, \psi^{(+)}_{im},
\psi^{(-)}_{im} ,\psi^{(+)}_s, \psi^{(-)}_s]^T={\bf C}^{-1}\cdot{\bf h}$ are the eigenfunctions of the counter-propagating (denoted by `-' and `+') magnetic mode ($\psi^{(\pm)}_m$), IMWs ($\psi^{(\pm)}_{im}$) and SDWs ($\psi^{(\pm)}_{s}$), respectively. Therefore, the matrix ${\bf C}$ relating the fluid variables  $\bf h$ to the eigenfunctions $\bf \psi$ is referred to as the transformation matrix. Because ${\bf \Lambda}$ is diagonal, these eigenfunctions are dynamically decoupled and evolve independently from each other with their corresponding eigenfrequencies,
\[
\psi^{(\pm)}_m\propto e^{\pm {\rm i}\int^t\omega_m(t') dt'},
\]
\[
\psi^{(\pm)}_{im}\propto e^{\pm {\rm i}\int^t\omega_{im}(t') dt'},
\]
\[
\psi^{(\pm)}_s\propto e^{\pm {\rm i}\int^t\omega_s(t') dt'}.
\]
The absence of coupling among the modal equations (\ref{eq:modal_eqs}) means that in the WKB regime, the modes evolve without exchanging energy among each other, i.e., initially exciting one mode with a specific time-scale of variation does not lead to the excitation of other modes with different time-scales. We will see below that in the non-WKB, or nonadiabatic regime, these modal equations become coupled due to shear, resulting in the dynamical coupling of the modes.

\begin{figure*}
\includegraphics[scale=0.32]{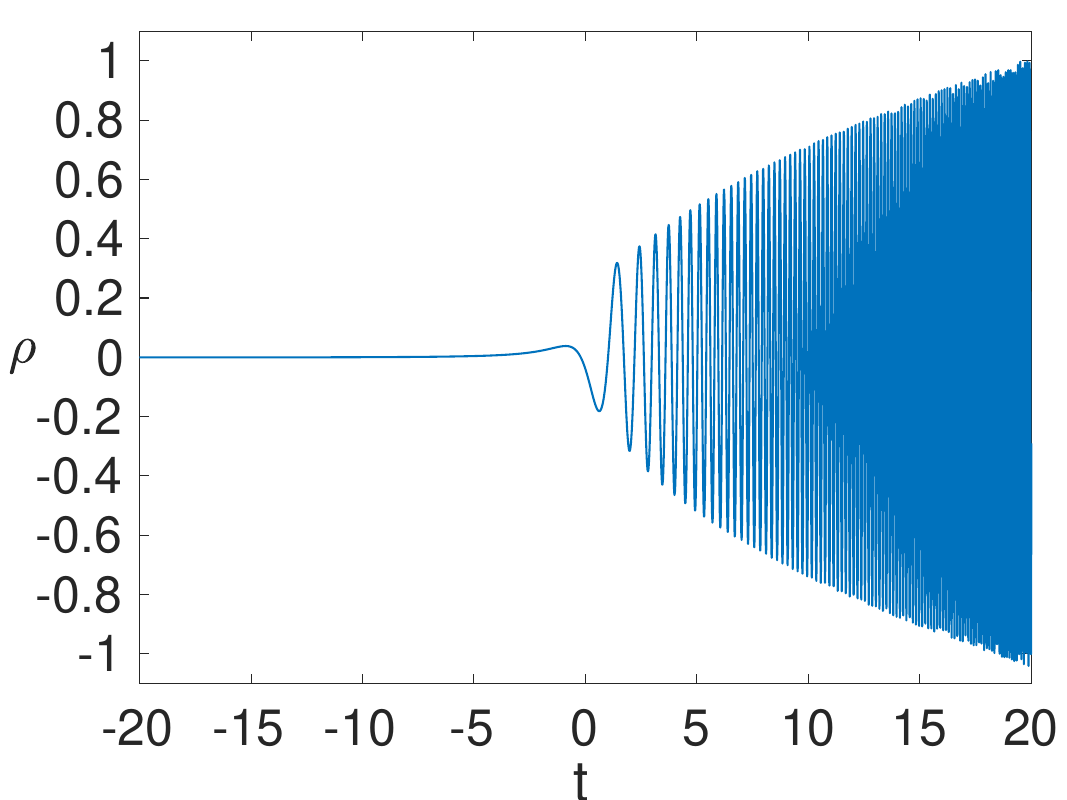} 
\includegraphics[scale=0.32]{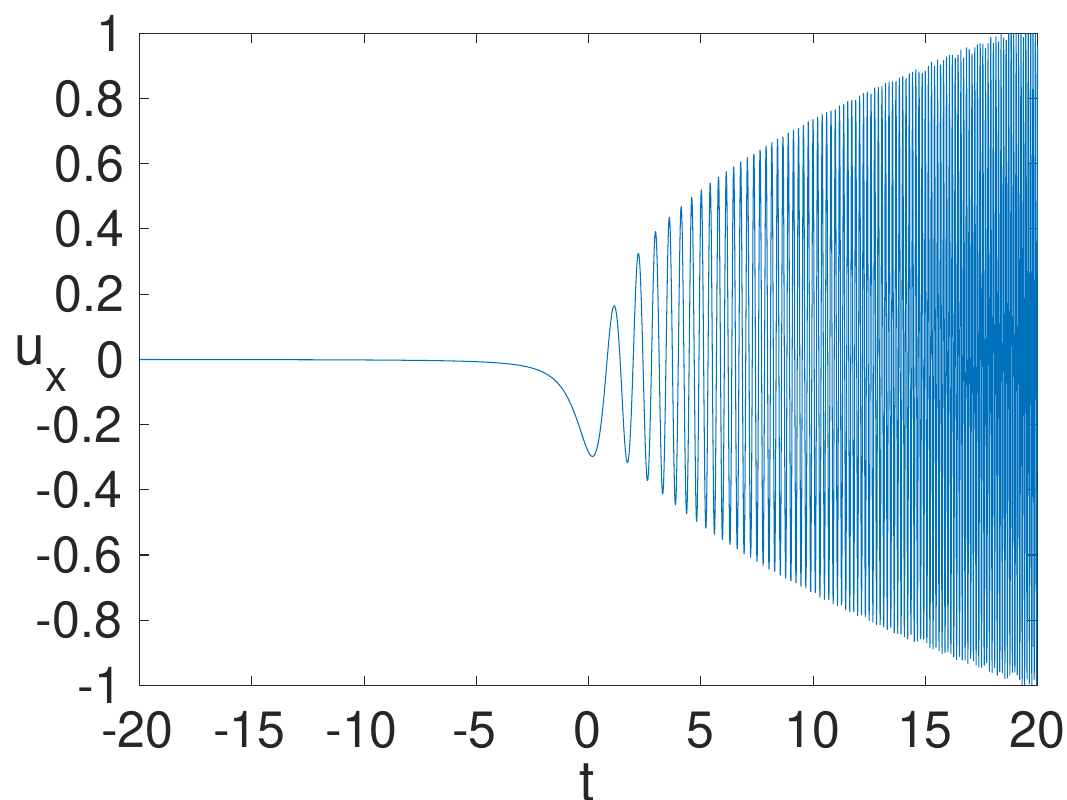}
\includegraphics[scale=0.32]{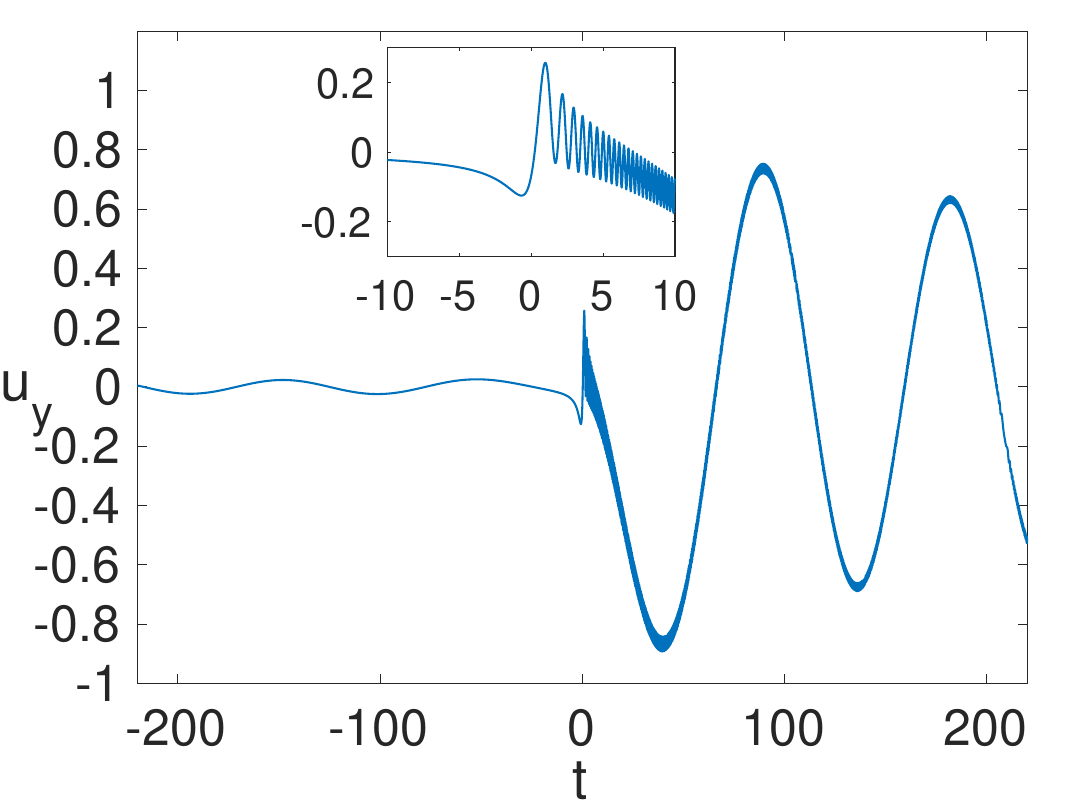} 
\includegraphics[scale=0.32]{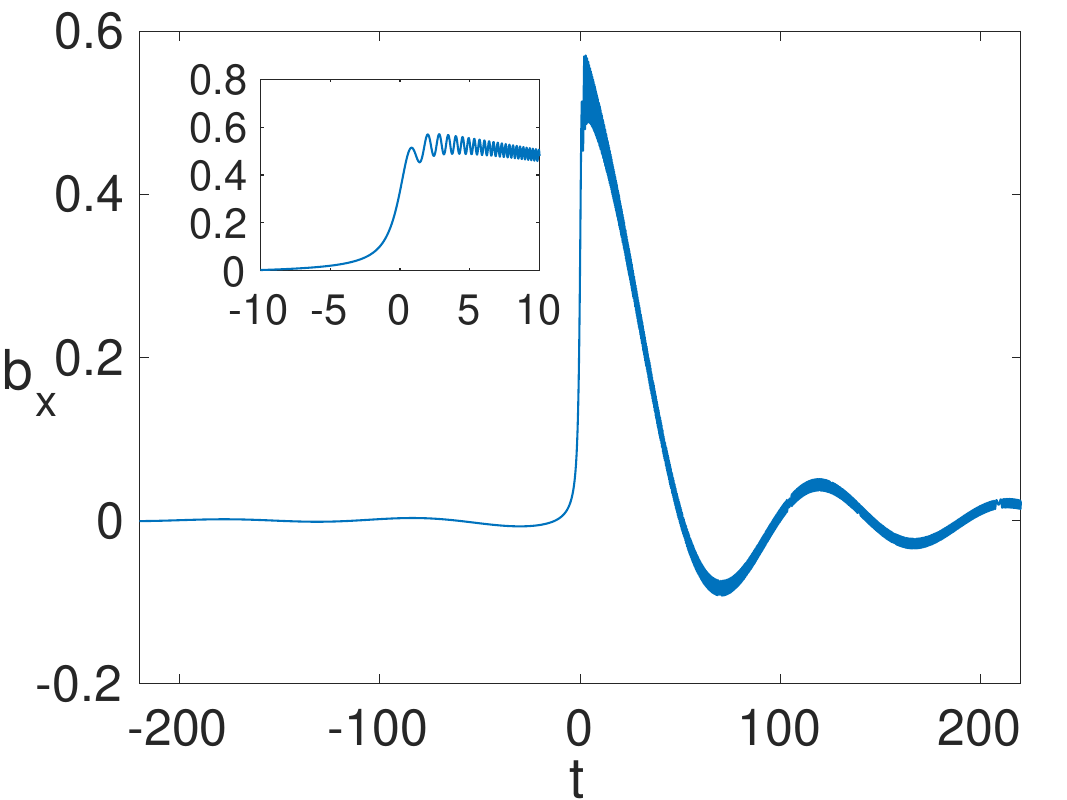}
\includegraphics[scale=0.32]{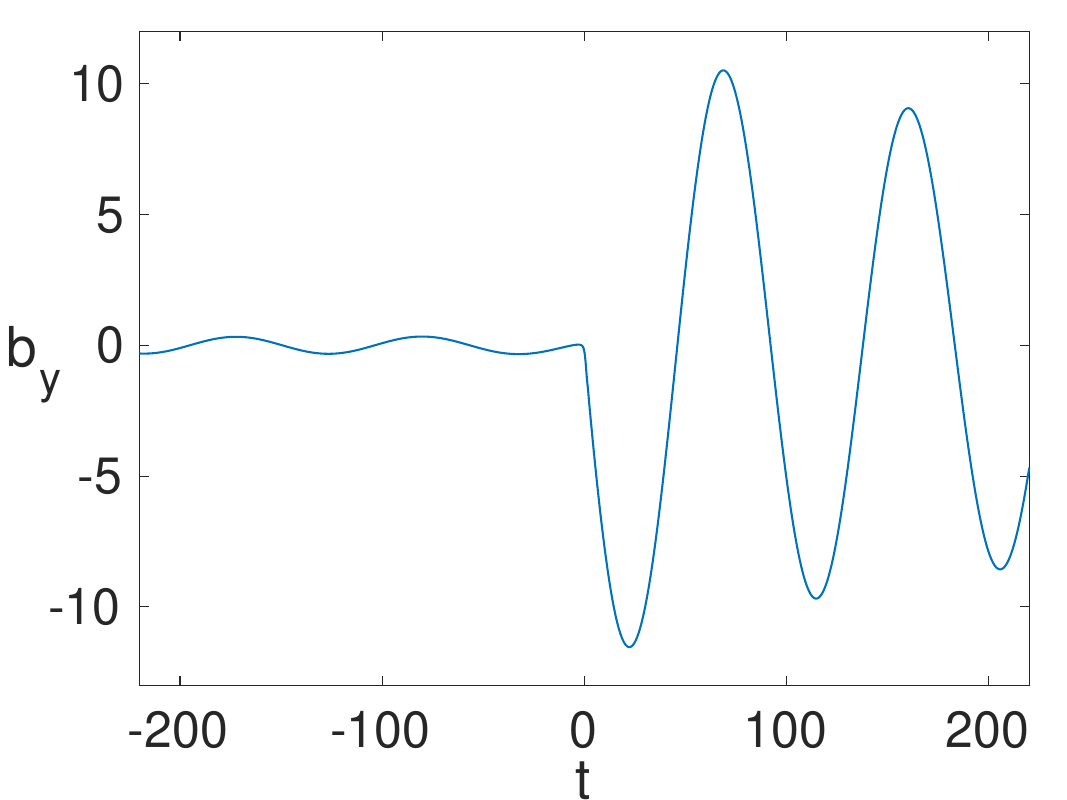} 
\includegraphics[scale=0.32]{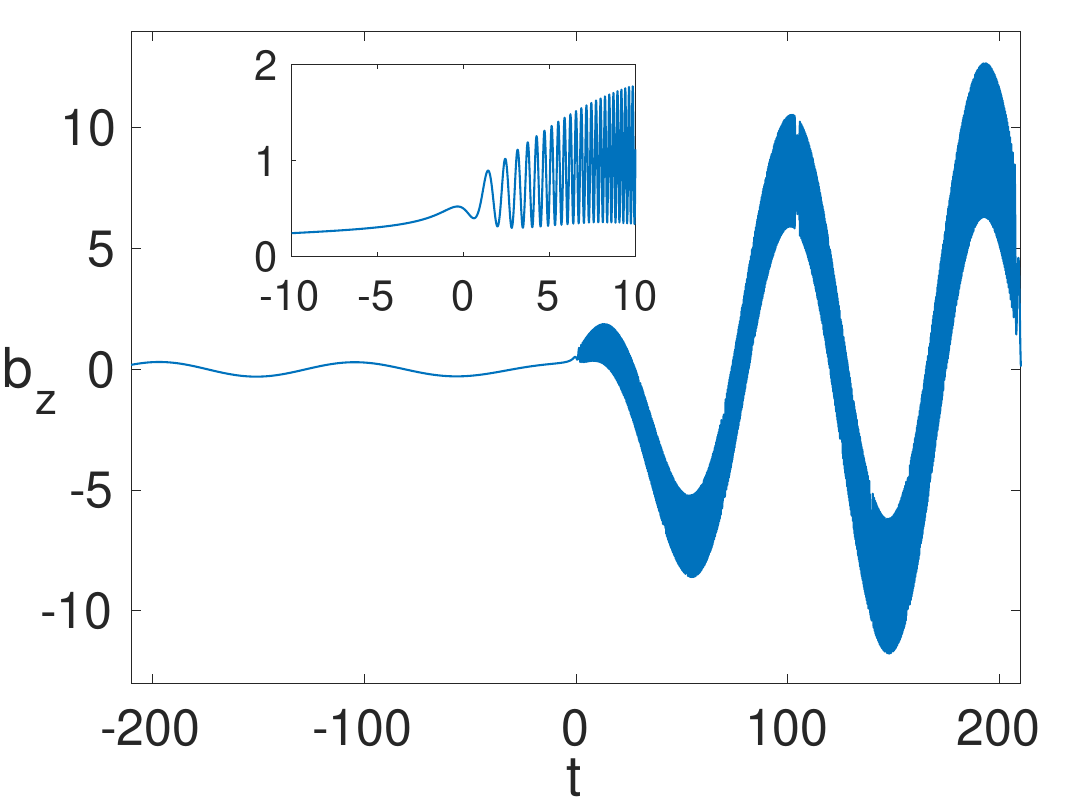}
\caption{
Evolution of the real parts of the density, velocity and magnetic field components pertaining to the initially imposed magnetic mode shearing wave with $k_y=2$ and $k_z=1$. The dominant components in the early evolution of the magnetic mode are transverse $b_y$ and $b_z$, which oscillate with lower frequency $\omega_m$ close to  the Alfv\'en frequency $v_Ak_z$, while longitudinal components $b_x$ are $u_x$ are almost zero. Then, $k_x$ enters the MRI-active range (\ref{eq:condMRI}), where all three field components undergo nonmodal MRI growth. When $k_x$ crosses the point $k_x=0$ at $t=0$, high-frequency oscillations abruptly emerge, which are strongest in $\rho$ and $u_x$ but also noticeable in $b_x$, $u_y$, $b_y$ and $b_z$. After leaving the MRI-active interval, $u_y, b_y$ and $b_z$, being mainly related to the original magnetic mode and generated IMWs, continue to oscillate with the lower-frequency and much higher amplitude at large times. The excited fast oscillations have high frequency $\omega_s$ and indicate the excitation of SDWs.}\label{fig:evolution}
\end{figure*}

\subsection{The modal equations in the non-WKB regime -- the linear mode coupling}

In the WKB regime considered above, the time variation of $k_x(t)$ and hence of the mode eigenfrequencies have been neglected. However, at higher $k_y\gtrsim k_z$, when $k_x(t)$ of non-axisymmetric modes, drifting along the $k_x-$axis, passes through the interval $|k_x| \lesssim k_y$, the WKB regime breaks down, $|d\omega(t)/dt|\sim \omega^2(t)$, and one should take into account the time-dependence of $k_x(t)$ in the main equations (\ref{eq:rhokk})-(\ref{eq:bykk}), or equivalently, in the ${\bf A}$ matrix in equation (\ref{eq:matrix_eqs}). This
non-adiabatic regime is more important, since the shear-induced nonmodal effects manifest themselves in the mode dynamics in this regime. We now generalize the decoupled modal equations (\ref{eq:modal_eqs}) to the non-WKB regime. In this case, equation (\ref{eq:matrix_eqs}) has the same form
\begin{equation}
\frac{d{\bf h}}{dt}={\bf A}(t)\cdot{\bf h}, \label{eq:matrix_eqs_time}
\end{equation}
but ${\bf A}$ now depends on time through $k_x(t)=q\Omega k_yt$. Taking into account that the transformation matrix ${\bf C}$ also varies with time through $k_x(t)$, from equation (\ref{eq:matrix_eqs_time}) we obtain a more general system of linear modal equations both in the adiabatic and non-adiabatic regimes
\begin{equation}
\frac{d{\boldsymbol \psi}}{dt}=[{\bf \Lambda}(t)+{\bf
M}(t)]\cdot{\boldsymbol \psi}, \label{eq:modal_eqs_time}
\end{equation}
where the generalized eigenfunctions are given, as before, by
${\boldsymbol \psi}(t)={\bf C}^{-1}(t)\cdot{\bf h}(t)$. The eigenvalue matrix ${\bf \Lambda}(t)$ has
also become time-dependent through $k_x(t)$. The matrix
\begin{equation}\label{eq:M}
{\bf M}(t)=-{\bf C}^{-1}(t)\frac{d}{dt}{\bf C}(t)
\end{equation}
is a new term compared to modal equations (\ref{eq:modal_eqs}), which arises from the time-dependence of $k_x(t)$ and hence from shear $q$, because ${\bf M} \propto dk_x/dt=qk_y$, as follows from expression (\ref{eq:M}). In other words, this new matrix term is determined by shear $q$ and is important for non-axisymmetric modes.  To better see its role  in the mode dynamics, it is convenient to write equation (\ref{eq:modal_eqs_time}) componentwise
\begin{multline}
\frac{d\psi^{(+)}_m}{dt}-({\rm i}\omega_m+M_{11})\psi^{(+)}_m\\
=M_{12}\psi^{(-)}_m+M_{13}\psi^{(+)}_{im}+M_{14}\psi^{(-)}_{im}+M_{15}\psi^{(+)}_s+M_{16}\psi^{(-)}_s, \label{eq:modal_eqs_time_mp}
\end{multline}
\begin{multline}
\frac{d\psi^{(-)}_m}{dt}+({\rm i}\omega_m-M_{22})\psi^{(-)}_m\\
=M_{21}\psi^{(+)}_m+M_{23}\psi^{(+)}_{im}+M_{24}\psi^{(-)}_{im}+M_{25}\psi^{(+)}_s+M_{26}\psi^{(-)}_s, \label{eq:modal_eqs_time_mm}
\end{multline}
\begin{multline}
\frac{d\psi^{(+)}_{im}}{dt}-({\rm i}\omega_{im}+M_{33})\psi^{(+)}_{im}\\=M_{31}\psi^{(+)}_m+M_{32}\psi^{(-)}_m+M_{34}\psi^{(-)}_{im}+M_{35}\psi^{(+)}_s+M_{36}\psi^{(-)}_s, \label{eq:modal_eqs_time_imp}
\end{multline}
\begin{multline}
\frac{d\psi^{(-)}_{im}}{dt}+({\rm i}\omega_{im}-M_{44})\psi^{(-)}_{im} \\=
M_{41}\psi^{(+)}_m+M_{42}\psi^{(-)}_m+M_{43}\psi^{(+)}_{im}+M_{45}\psi^{(+)}_s+M_{46}\psi^{(-)}_s, \label{eq:modal_eqs_time_imm}
\end{multline}
\begin{multline}
\frac{d\psi^{(+)}_s}{dt}-({\rm i}\omega_s+M_{55})\psi^{(+)}_s=\\
=M_{51}\psi^{(+)}_m+M_{52}\psi^{(-)}_m+M_{53}\psi^{(+)}_{im}+M_{54}\psi^{(-)}_{im}+M_{56}\psi^{(-)}_s, \label{eq:modal_eqs_time_sp}
\end{multline}
\begin{multline}
\frac{d\psi^{(-)}_s}{dt}+({\rm i}\omega_s-M_{66})\psi^{(-)}_s=\\
=M_{61}\psi^{(+)}_m+M_{62}\psi^{(-)}_m+M_{63}\psi^{(+)}_{im}+M_{64}\psi^{(-)}_{im}+M_{65}\psi^{(+)}_s. \label{eq:modal_eqs_time_sm}
\end{multline}
It is evident that the matrix ${\bf M}$ couples the mode eigenfunctions as well as modifies their frequencies,
so we appropriately refer to it as a coupling matrix. All its
elements, being proportional to $q$ and $k_y$, are thus appreciable for non-axisymmetric modes and vanish in the WKB regime at large $k$ and small $k_y \ll |k_x|, k_z$. The latter condition includes axisymmetric modes with $k_y=0$, which are therefore not coupled in the linear regime. The coupling matrix $\bf M$ therefore encapsulates all shear-induced nonmodal effects. The left-hand sides of equations (\ref{eq:modal_eqs_time_mp})-(\ref{eq:modal_eqs_time_sm}) describe the individual dynamics of the magnetic mode, IMWs and SDWs, respectively, modified by shear via the diagonal components of ${\bf M}$ that give rise to the shear-induced nonmodal growth of these modes. For the magnetic mode, MRI remains the dominant amplification mechanism, while nonmodal effects, mediated by $M_{11}$ and $M_{22}$, further modify its dynamics. On the other hand, the right-hand sides, originating from the non-diagonal elements of ${\bf M}$, describe another main nonmodal effect due to shear -- mutual couplings among these modes as well as the interaction between counter-propagating components of each mode. Specifically, the non-diagonal elements
$M_{12}$ and $M_{21}$, $M_{34}$ and $M_{43}$, $M_{56}$ and $M_{65}$
describe the coupling, respectively, between the counter-propagating
components, $\psi^{(+)}_m$ and $\psi^{(-)}_m$, of the magnetic mode,
$\psi^{(+)}_{im}$ and $\psi^{(-)}_{im}$ of IMWs and
$\psi^{(+)}_s$ and $\psi^{(-)}_s$ of SDWs. The other non-diagonal
elements describe the couplings among different modes: $M_{13},M_{14},M_{23},M_{24}$ and
$M_{31},M_{41},M_{32},M_{42}$ describe the coupling between the magnetic mode and IMWs, $M_{53},M_{54},M_{63},M_{64}$ and
$M_{35},M_{45},M_{36},M_{56}$ describe the coupling between IMWs and SDWs and $M_{51},M_{52},M_{61},M_{62}$ and
$M_{15},M_{25},M_{16},M_{26}$ describe the coupling between the
SDWs and the magnetic mode. In other words, due to  these coupling terms, an initially imposed one of these modes can act as a source for the other modes and excite them during evolution.

All these non-diagonal components reach their maximum and are appreciable in the non-adiabatic range $|k_x(t)|\lsim |k_y|$, where the nonmodal effects are important  
\citep{Tevzadze_etal10, Mamatsashvili_Rice2011, Mamatsashvili_etal2013}. In our previous paper \cite{Mamatsashvili_etal2013}, we analyzed the nonmodal growth and coupling of inertia-gravity waves and magnetic mode in incompressible magnetized discs. Although the system of coupled equations (\ref{eq:modal_eqs_time_mp})-(\ref{eq:modal_eqs_time_sm}) is quite complex involving all possible mode couplings, here we focus specifically on the coupling between SDWs and the dominant, MRI-unstable magnetic mode, which is determined by the coefficients $M_{51},M_{52},M_{61},M_{62}$ shown in Fig. \ref{fig:coupling_coeff}. They are appreciable in the non-adiabatic range
$|k_x(t)|\lsim k_y$, peaking around $k_x=0$, and indicate that the shear-induced coupling between these modes  occurs just in this  range. At large $|k_x(t)|\gg k_y,k_z$, in the WKB regime, the coupling matrix vanishes, ${\bf M}\rightarrow 0$, and hence  there is no energy
exchange among the modes. \footnote{In fact, other elements of the $\bf M$ matrix behave similarly with $k_x$, so we do not show them here.} Below, we characterize this SDW-MRI coupling in more detail.

\begin{figure*}
\includegraphics[scale=0.32]{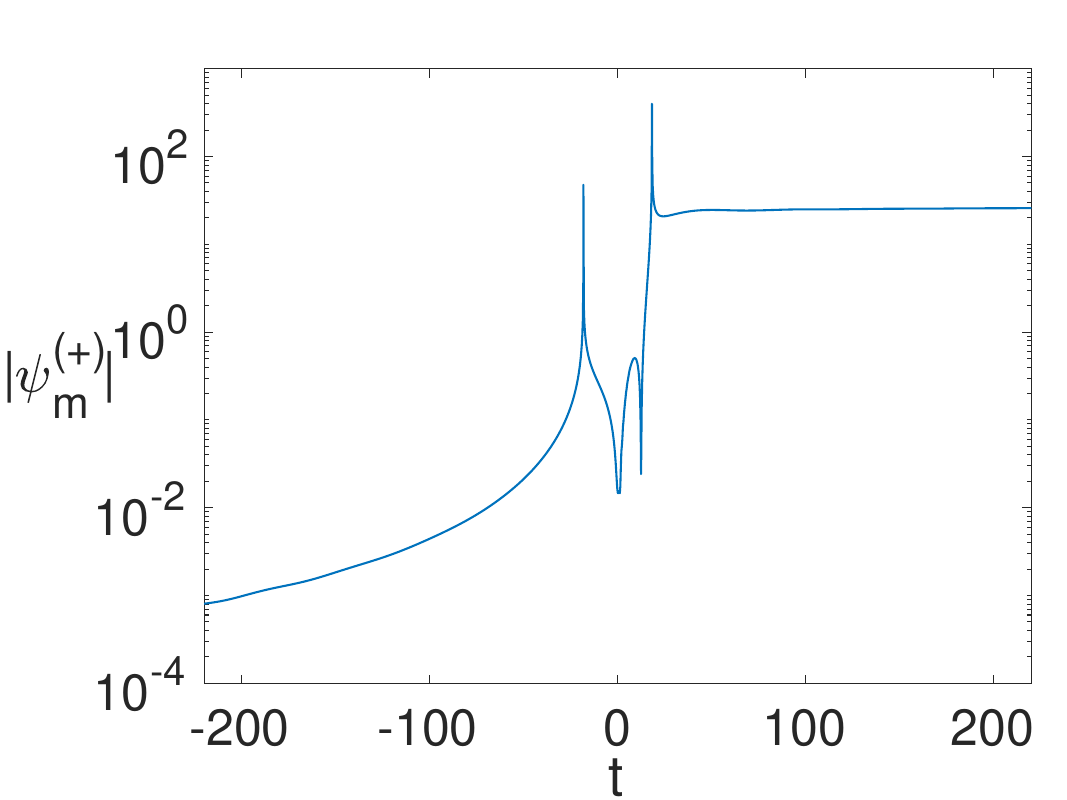} 
\includegraphics[scale=0.32]{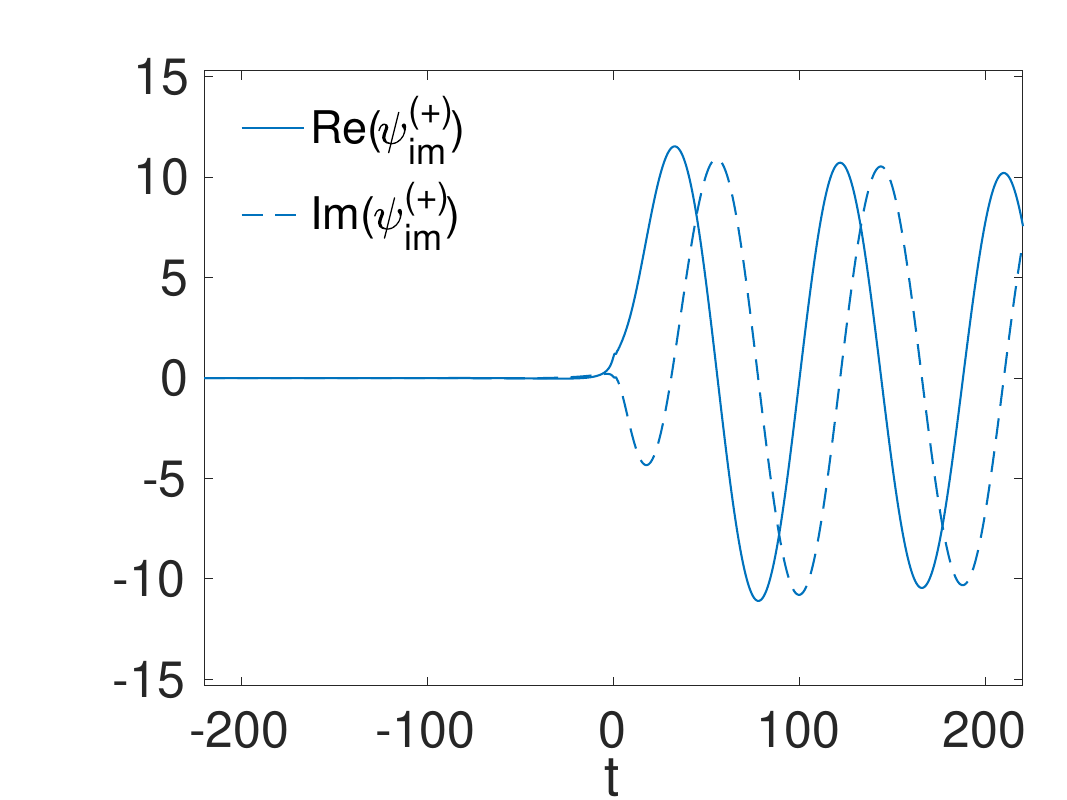}
\includegraphics[scale=0.32]{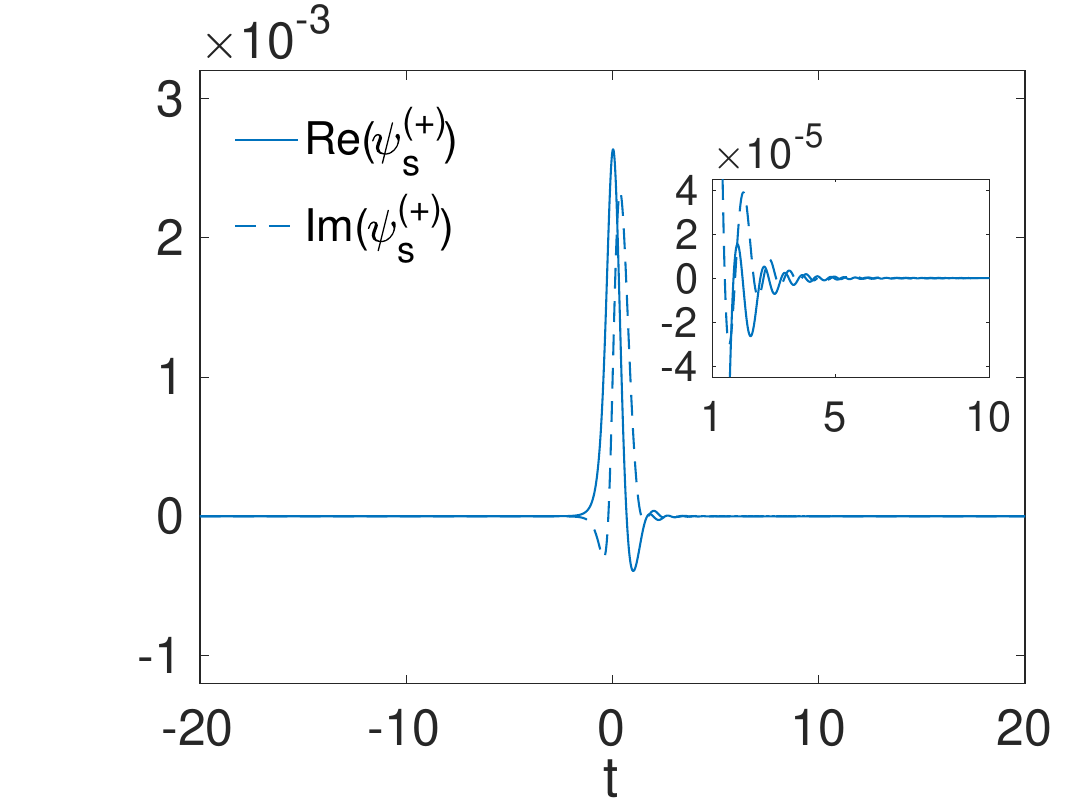} 
\includegraphics[scale=0.32]{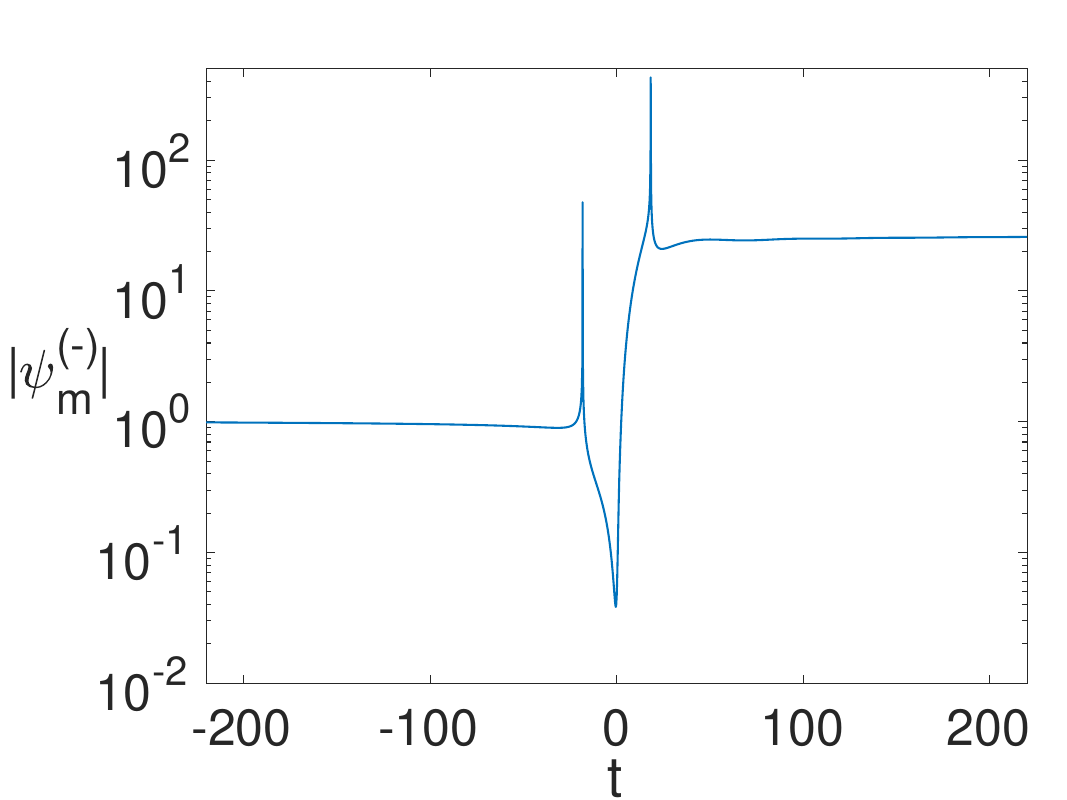}
\includegraphics[scale=0.32]{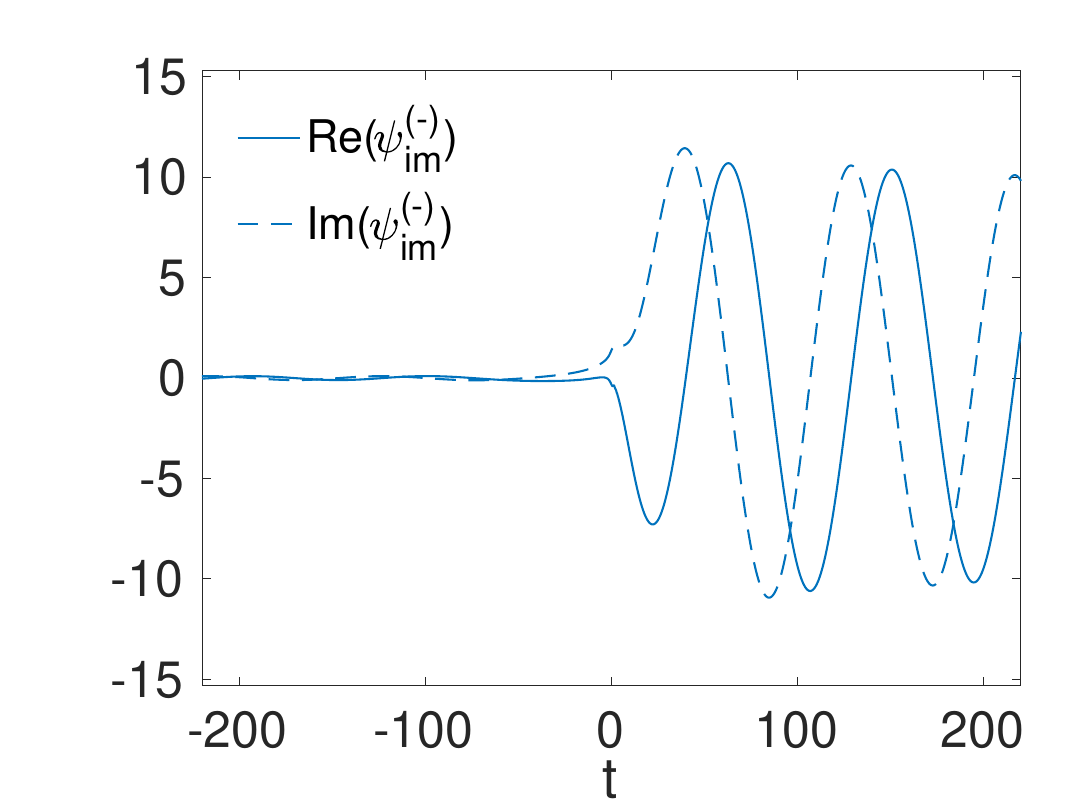} 
\includegraphics[scale=0.32]{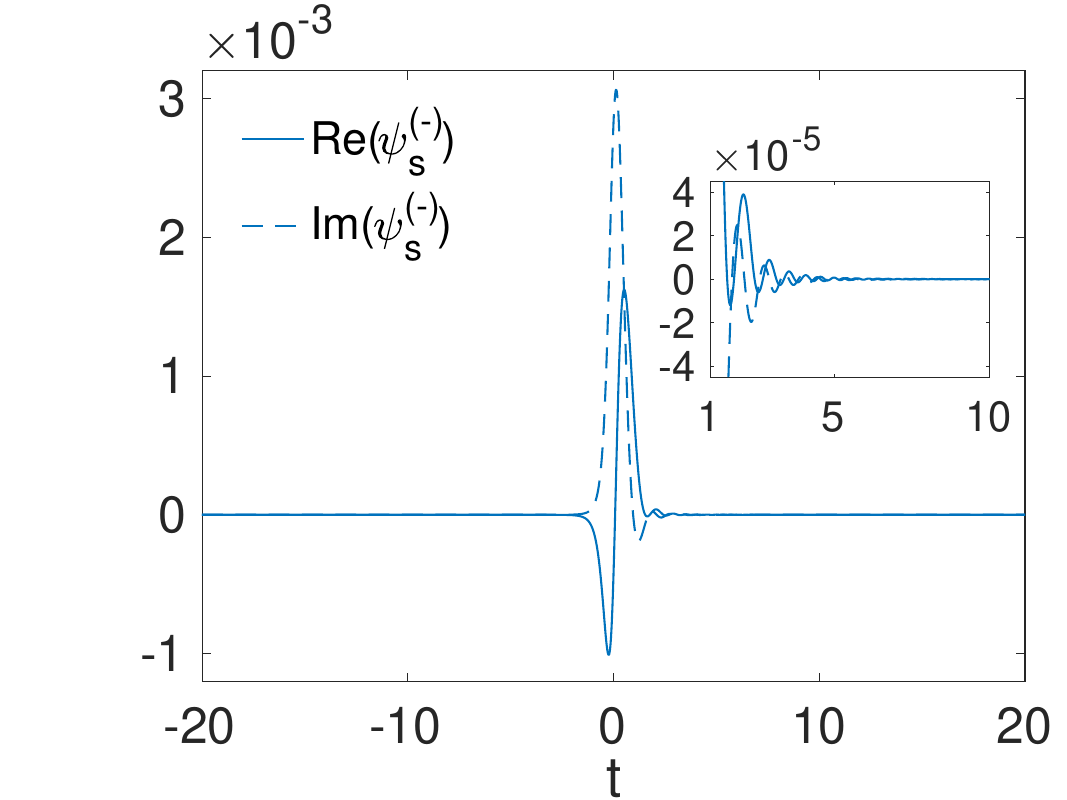}
\caption{Evolution of the eigenfunctions at $k_y=2$ and $k_z=1$, which correspond to the perturbations in Fig. \ref{fig:evolution}, for the initially imposed magnetic mode $\psi_m^{(-)}$. $\psi_m^{(-)}$, undergoing nonmodal MRI growth in the instability region $|t|\leq 18.3$ (see text), induces the other component $\psi_m^{(+)}$ via the coupling term $M_{12}$. The latter component in turn, feeding back on $\psi_m^{(-)}$ via the coupling term $M_{21}$, causes its decrease around $t=0$ over a short period of time. Then, at $t>0$ both $\psi_m^{(\pm)}$ increase (to highlight their growth, only the absolute values of $\psi_m^{(\pm)}$ are shown). On the other hand, $\psi_{im}^{(\pm)}$ and $\psi_{s}^{(\pm)}$ remain nearly zero at $t<0$ and at about  $t=0$ display emergence of low-frequency ($\omega_{im}$) and high-frequency ($\omega_s$) oscillations, indicating the excitation of IMWs and SDWs, respectively.} \label{fig:eigenfunctions}
\end{figure*}

\section{Evolution of the modes -- coupling of MRI and SDWs}

In this section, we investigate the dynamics of the non-axisymmetric magnetic mode, its nonmodal MRI growth and generation of SDWs by numerically solving system (\ref{eq:matrix_eqs_time}). At large times $|t|\gg 1$, when $|k_x(t)|\gg k_y,k_z$, as discussed above, the WKB condition holds and the coupling matrix vanishes, implying that the modes are dynamically decoupled from each other. This allows us to impose only the magnetic mode in the beginning, as an initial condition for equations (\ref{eq:matrix_eqs_time}), and follow its dynamics. 

To prepare such initial conditions, initially at
$t=-t_0$, where $t_0\gg 1$ is some large parameter, we impose a tightly leading ($k_x(-t_0)/k_y<0, |k_x(-t_0)|/k_y\gg 1$) shearing wave of the purely magnetic mode and trace subsequent evolution until $t=t_0$, when shearing waves of the modes become tightly trailing ($k_x(t_0)/k_y \gg 1$). Without loss of generality, from the two counter-propagating components of the magnetic mode, we choose the one given by
$\psi^{(-)}_m$ as an initial condition at $t=-t_0$. In this
adiabatic regime, ignoring the mode coupling, this eigenfunction is given by the asymptotic WKB solution of the homogeneous, left hand side, part of equation (\ref{eq:modal_eqs_time_mm}),
\begin{equation}\label{eq:init_psi_m}
\psi_m^{(-)}=C_0e^{-\int_{-t_0}^t [{\rm i}{\omega}_m(t')-M_{22}(t')]dt'}=e^{-{\rm i}v_Ak_zt}
\end{equation}
where $C_0$ is some arbitrary constant setting the initial value, which is chosen such that at $t=-t_0$, $\psi_m^{(-)}$ has a unit amplitude and oscillates with the Alfv\'en frequency $v_Ak_z$, since $\omega_m\rightarrow v_Ak_z$, at large $|k_x(-t_0)|$. The other component of the magnetic mode $\psi_m^{(+)}$ as well as IMWs and SDWs are not excited initially, that is, their corresponding eigenfunctions are set to zero in the beginning, $\psi_m^{(+)}(-t_0)=\psi_{im}^{(\pm)}(-t_0)=\psi_s^{(\pm)}(-t_0)=0$. These conditions ensure that initially only the magnetic mode is present, with a single direction of propagation, whereas IMWs and SDWs are absent. The corresponding initial state vector ${\bf h}_m$ associated with this magnetic mode is found as ${\bf h}_m(-t_0)={\bf
C}(-t_0)\cdot[0,\psi^{(-)}_m(-t_0),0,0,0,0]^{T}$, which is then inserted in equation (\ref{eq:matrix_eqs_time}) as an initial condition. For numerical integration we use a standard Runge-Kutta scheme (MATLAB ode45 RK implementation).

Figure \ref{fig:evolution} shows the subsequent evolution of the perturbed density, velocity and magnetic field components, while Fig. \ref{fig:eigenfunctions} shows the evolution of the corresponding mode eigenfunctions. In the beginning, being in the adiabatic region $|k_x(t)|\gg k_y$ and far from the coupling region, the initially imposed magnetic mode oscillates with the frequency $\omega_m$, which at this large $k_x$ is close to  the Alfv\'en frequency, $\omega_m\approx  k_z\sqrt{2/\beta}$, and induces similar oscillations in the velocity and magnetic field. The primary components in this mode are the azimuthal and vertical velocities, $u_y$ and $u_z$, and magnetic field $b_y$ and $b_z$, which oscillate with a nearly constant amplitude, whereas the density $\rho$ as well as the radial velocity $u_x$ and field $b_x$  remain nearly zero, since the magnetic mode itself is incompressible. 

The dominance of the magnetic mode and the absence of IMWs and SDWs during the early phase of the evolution are also evident from the  eigenfunctions in Fig. \ref{fig:eigenfunctions}.  Since the coupling matrix is small (Fig. \ref{fig:coupling_coeff}), $\psi_s^{(\pm)}$,  $\psi_{im}^{(\pm)}$ and the counter-propagating branch of the magnetic mode $\psi_{m}^{(+)}$ remain all very small compared to $\psi_{m}^{(-)}$, which oscillates with a constant amplitude according to equation (\ref{eq:init_psi_m}).

As time progresses, $k_x(t)$ of the magnetic mode approaches and begins to cross the region $|k_x(t)/k_y|\lsim 1$. In this region, the flow shear comes into play giving rise to two main effects: (i) the magnetic mode no longer oscillates but undergoes nonmodal MRI growth and (ii) the characteristic time-scales of the SDWs, IMWs, the MRI growth time and the shear time $1/q$ (in the non-dimensional form) all come close to each other, i.e., $\omega_s\sim \omega_{im}\sim \gamma_m\sim q$. Since the time-scales of the modes are comparable now, the coupling matrix elements become appreciable (Fig. \ref{fig:coupling_coeff}).  As a consequence, an efficient energy exchange -- nonmodal couplings -- among these  modes as well as between the magnetic mode and the disc flow via MRI occurs during this stage. Note also that, in this interval, the adiabatic condition for the mode frequencies breaks down, $|d\omega(t)/dt|\sim \omega^2(t)$, which is another reason for mode coupling \citep{Chagelishvili_etal96, Gogoberidze_etal2004} \footnote{Note that this is not the case for small $k_y\ll k_z$, when the dynamics is uniformly adiabatic at all times even at $|k_x(t)/k_y|\lsim 1$, as discussed in Section 2.2}. Let us look in more detail at the evolution of variables and mode interactions in this coupling range.

\begin{figure*}
\includegraphics[scale=0.32]{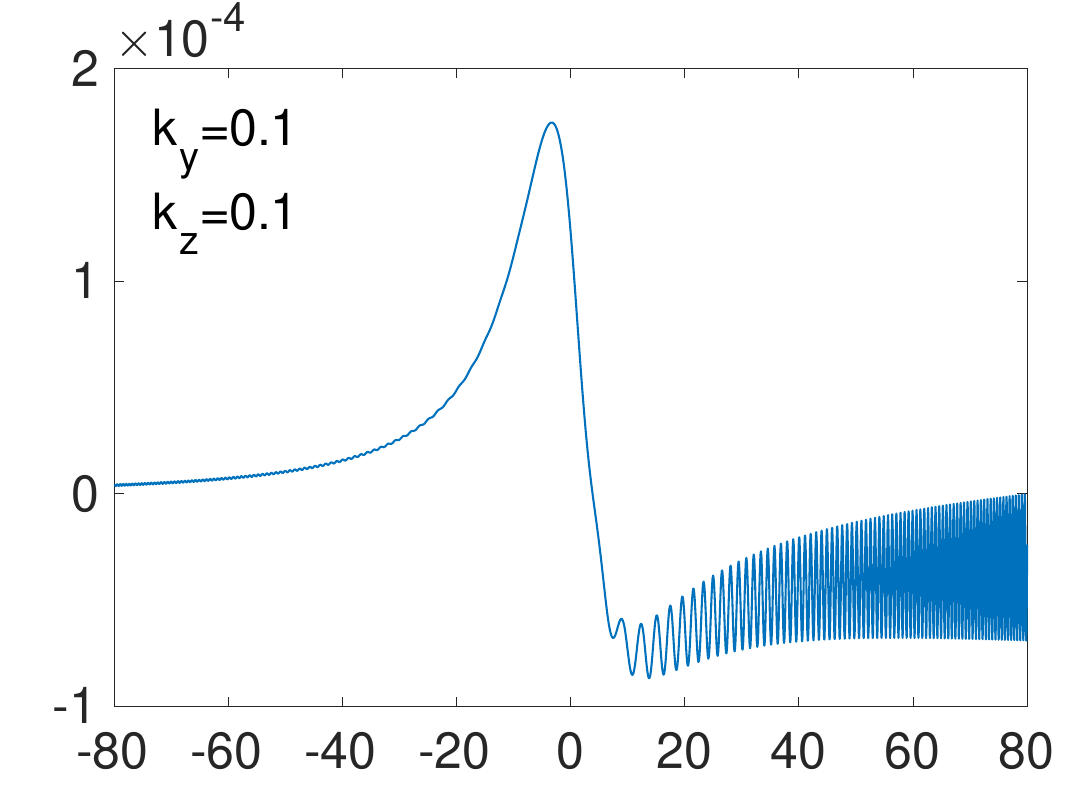} 
\includegraphics[scale=0.32]{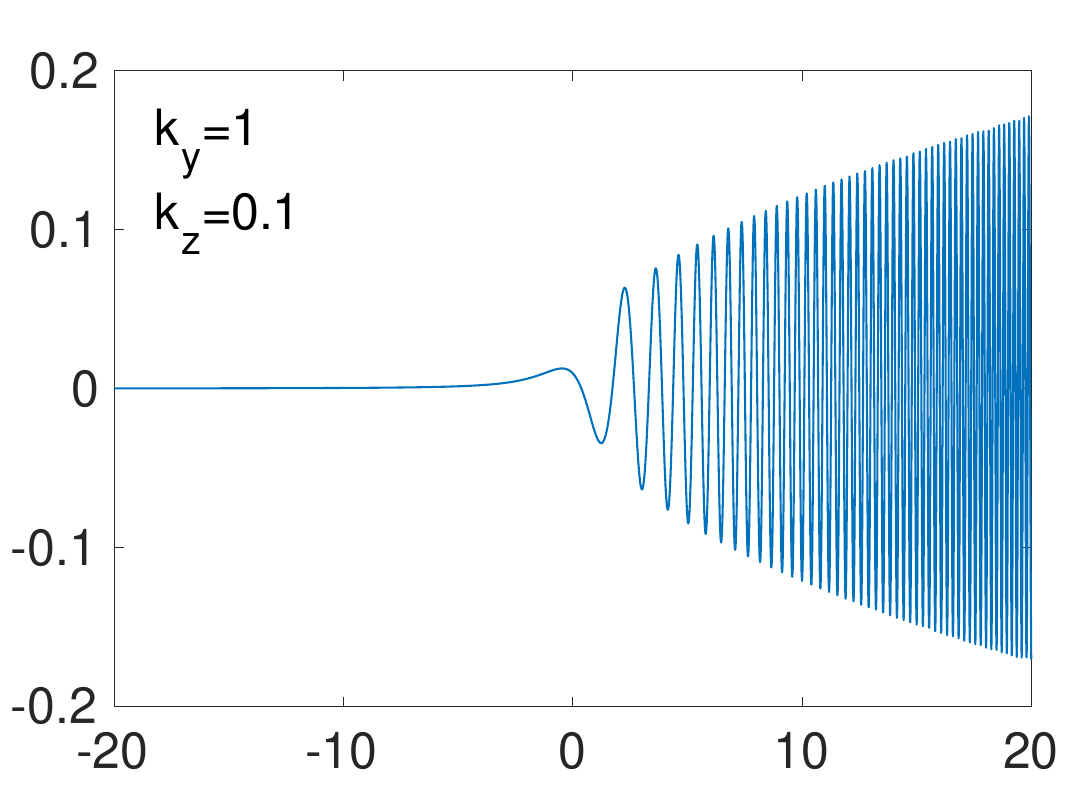}
\includegraphics[scale=0.32]{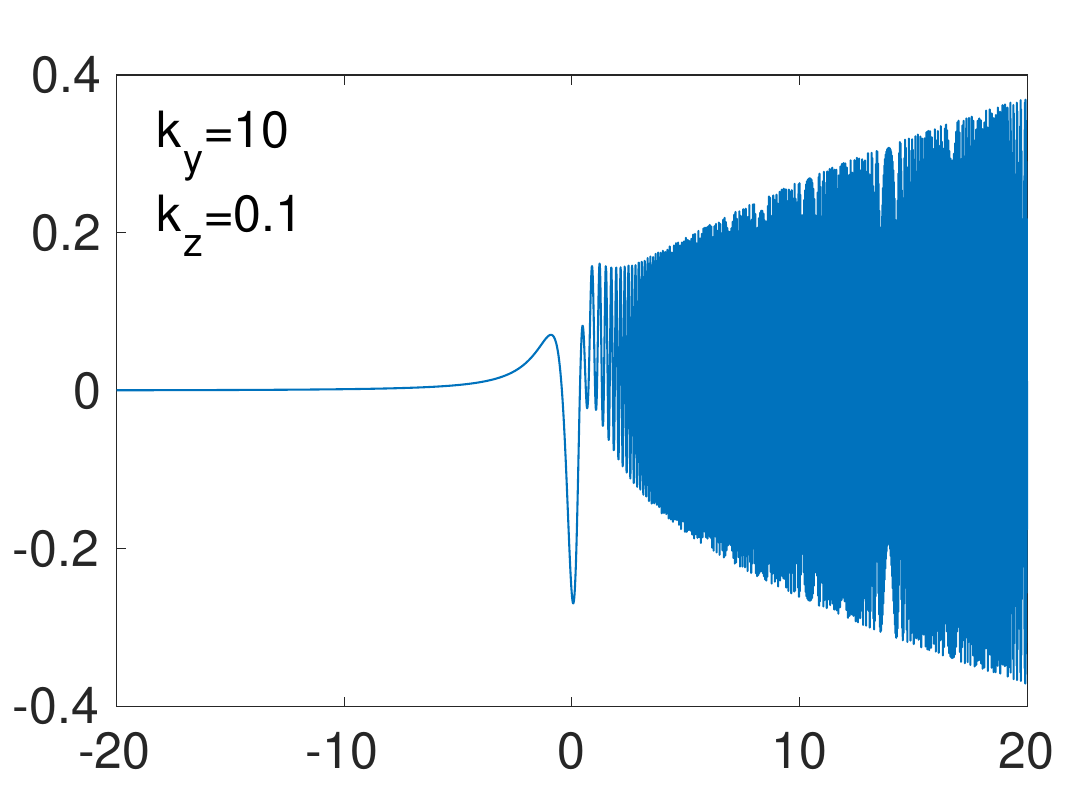} 
\includegraphics[scale=0.32]{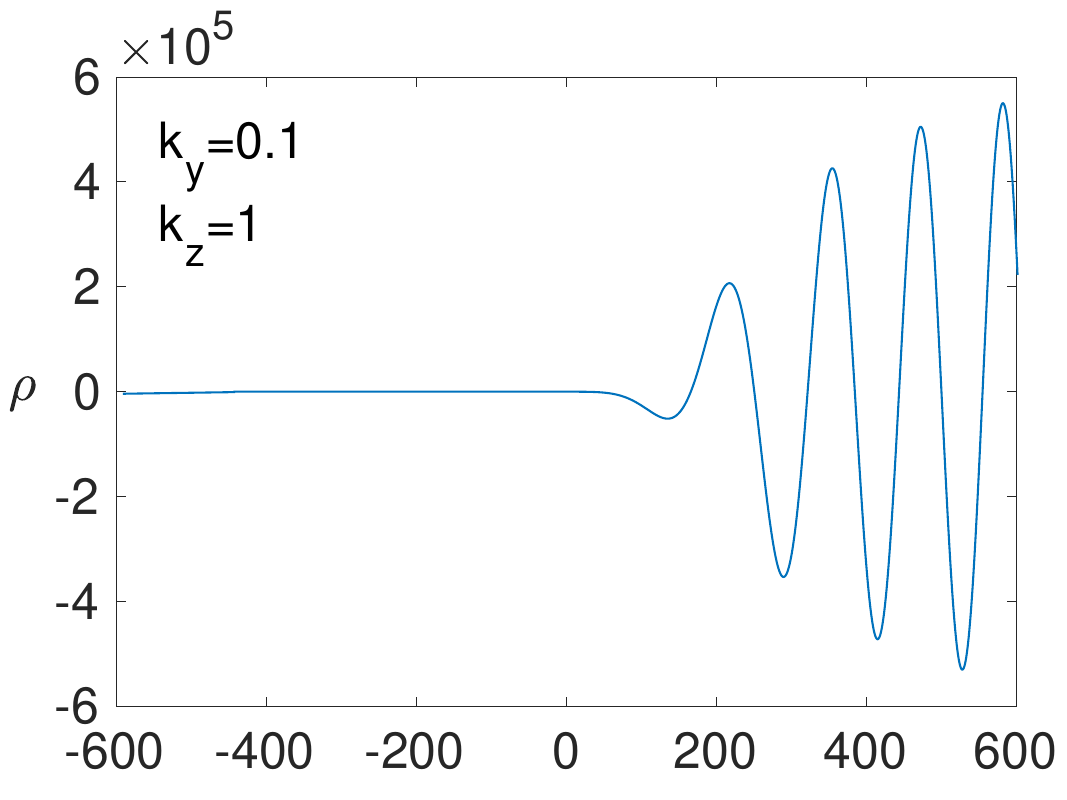} 
\includegraphics[scale=0.32]{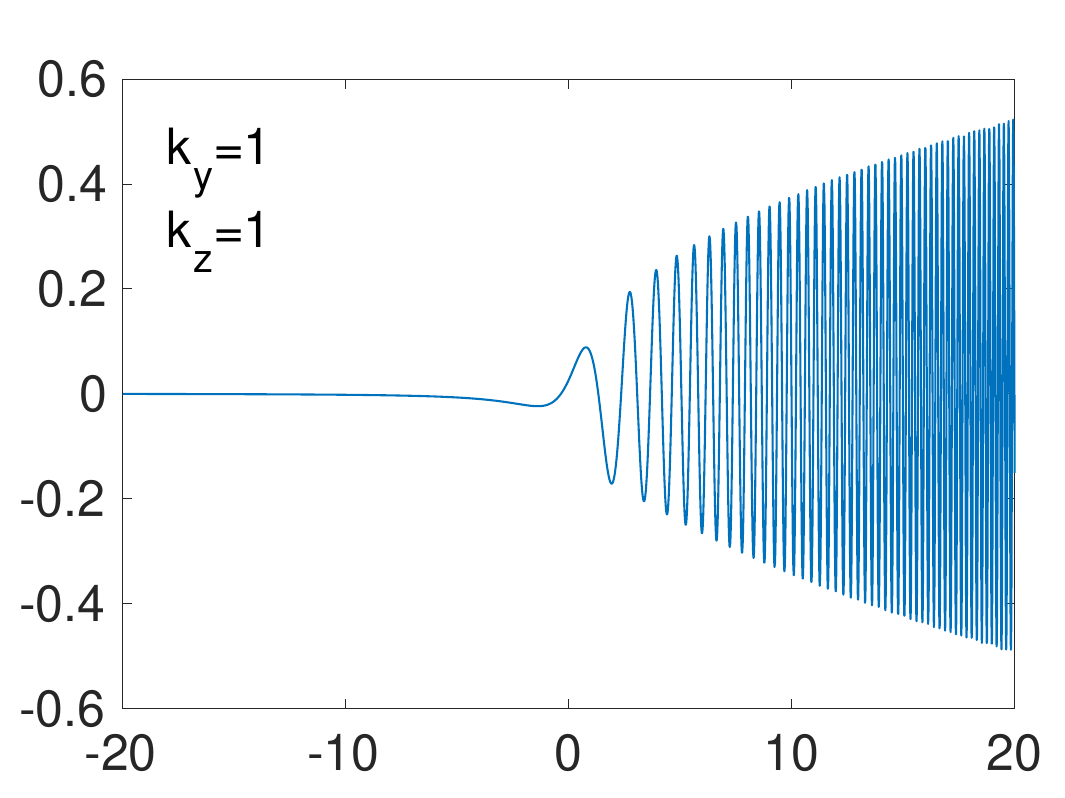}
\includegraphics[scale=0.32]{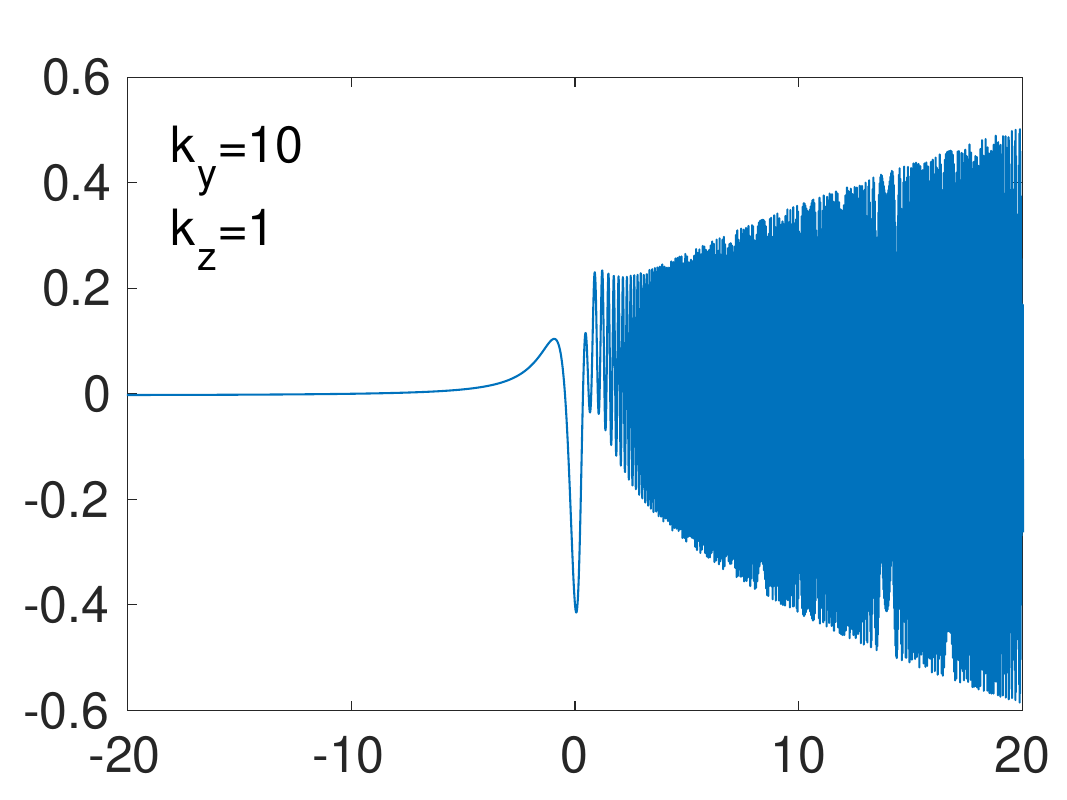} 
\includegraphics[scale=0.32]{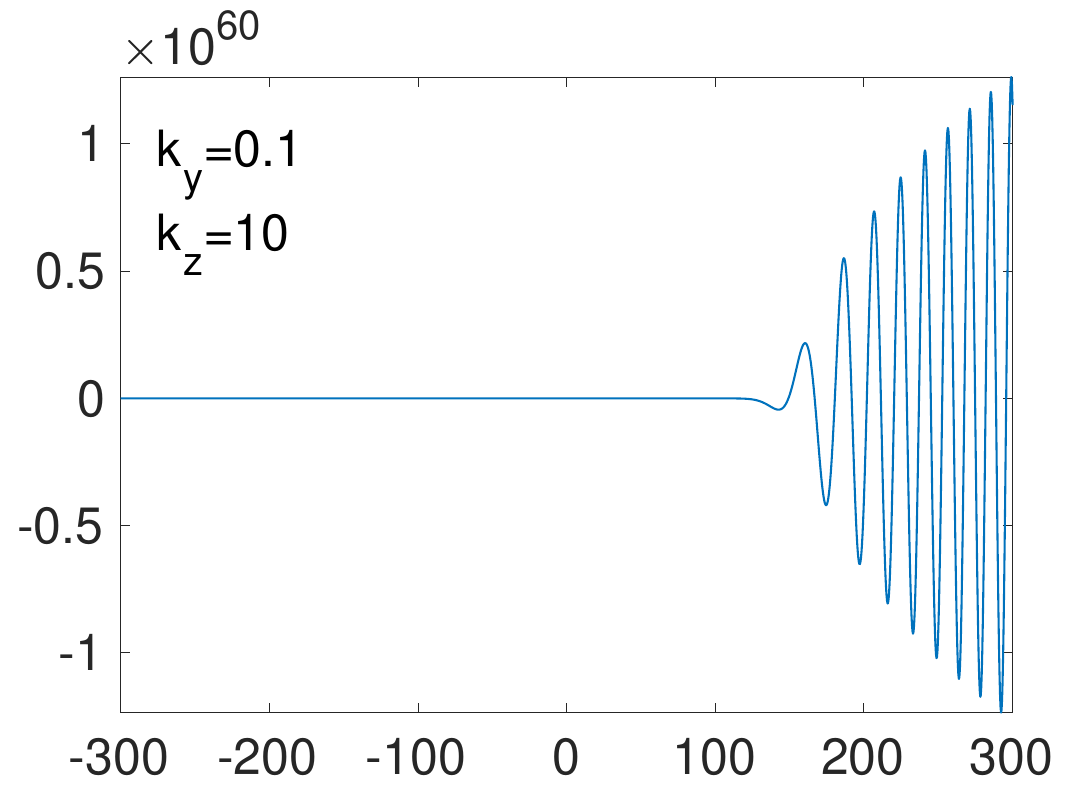} 
\includegraphics[scale=0.32]{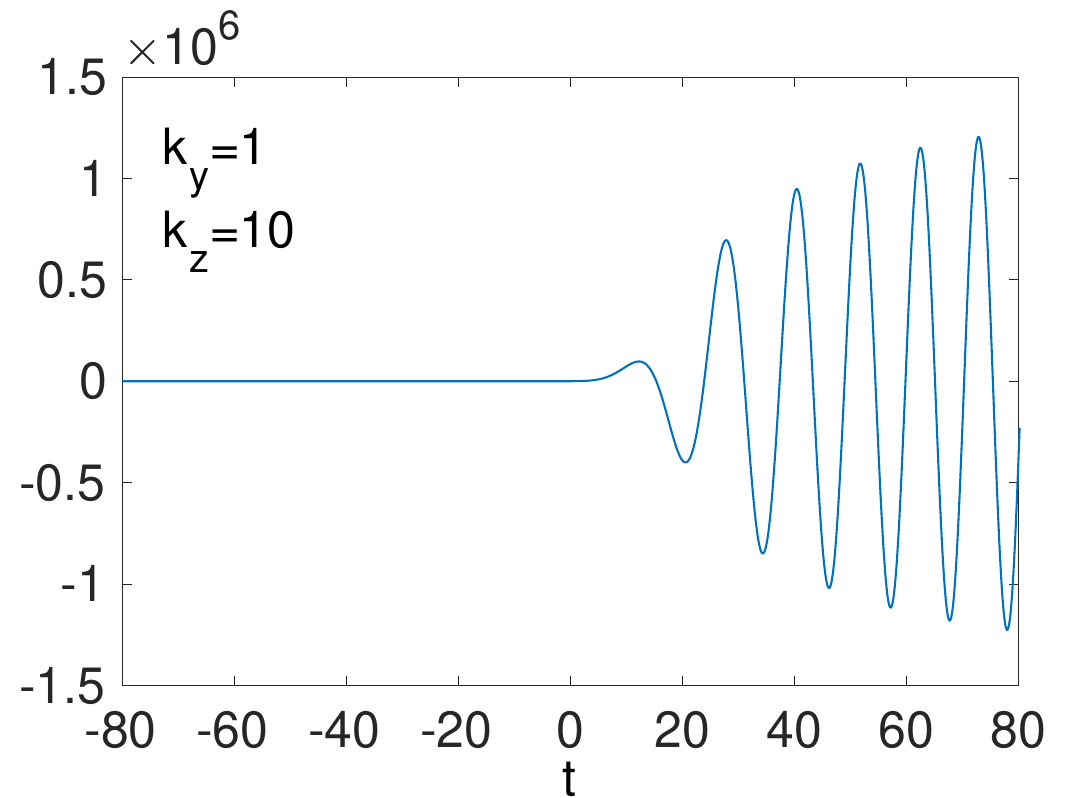}
\includegraphics[scale=0.32]{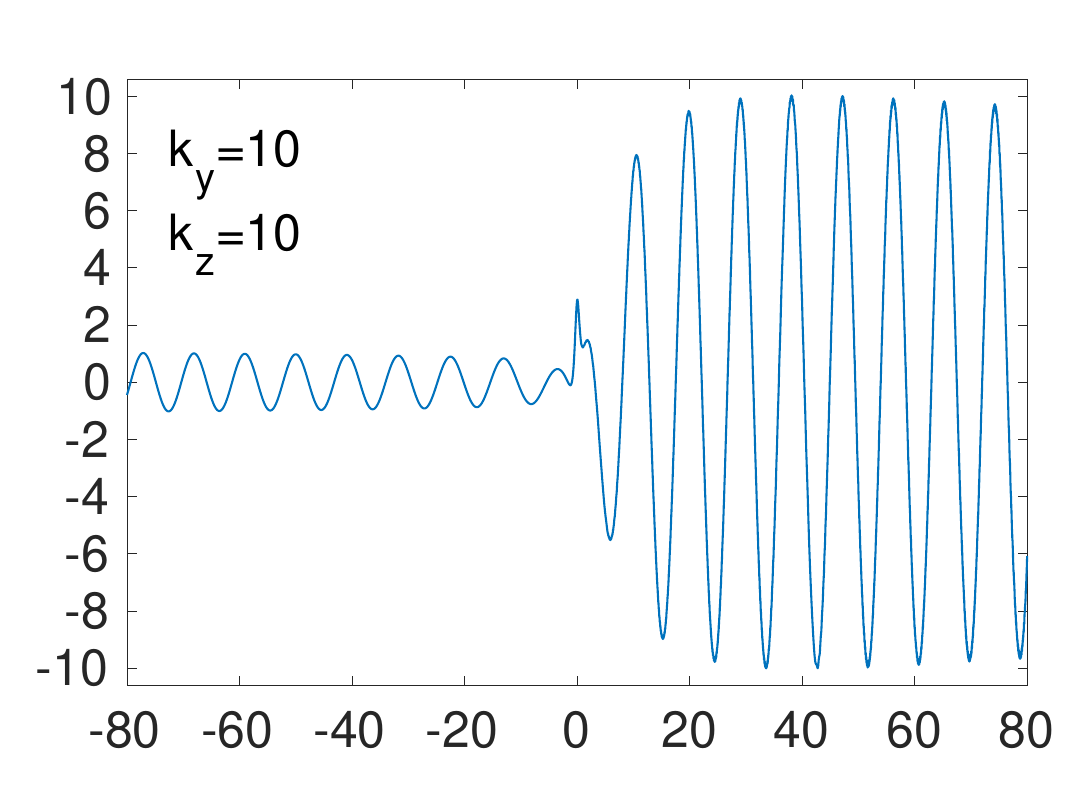} 
\caption{Evolution of density at different $k_y$ and $k_z$ for the same initial conditions consisting of a single magnetic mode component $\psi_m^{(-)}$. The SDW generation, indicated by high-frequency oscillations with increasing frequency and amplitude, tends to be more efficient at higher $k_y$ and lower $k_z$, i.e., $k_z\lesssim k_y$.}\label{fig:density_diffkykz}
\end{figure*}

First, at $t<0$, when $k_x(t)$ enters the MRI-active range, $|k_x|/k_y\leq [(q\beta/k_z^2-1)(1+k_z^2/k_y^2)]^{1/2}=27.4$, or $|t|\leq 18.3$ (equation \ref{eq:condMRI}), the magnetic mode becomes non-oscillatory and starts to grow due to MRI, as seen  in the evolution of $\psi_m^{(-)}$ in Fig. \ref{fig:eigenfunctions}.  The coupling term $M_{12}$ becomes appreciable in this interval, resulting in the generation of the other component,  $\psi_m^{(+)}$, of the magnetic mode by the initially imposed one $\psi_m^{(-)}$, as  also shown in Fig. \ref{fig:eigenfunctions}. The newly generated $\psi_m^{(+)}$ subsequently feeds back on the dynamics of $\psi_m^{(-)}$ through the coupling term $M_{21}$, which is also appreciable near $k_x(t)=0$, thereby causing $\psi_m^{(-)}$ to decrease around $t=0$. Afterwards both $\psi_m^{(-)}$ and $\psi_m^{(+)}$ grow due to MRI and mutual coupling, respectively, until they leave this MRI-active region.  MRI in turn leads to the growth of all the velocity and magnetic field components. Since the coupling coefficients are larger now, the growth of $\psi_m^{(-)}$ also induces a moderate amplification of $\psi_s^{(\pm)}$ and
$\psi_{im}^{(\pm)}$ due to the corresponding coupling terms. Then, at around $t=0$, when $k_x(t)$ crosses the point
$k_x=0$, becoming positive $k_x(t)>0$, qualitatively different behaviour emerges -- high-frequency oscillations abruptly appear in the evolution of $\psi_s^{(\pm)}$ and low-frequency oscillations in $\psi_{im}^{(\pm)}$, indicating the generation of rapidly varying SDWs and slowly varying IMWs, respectively (Fig. \ref{fig:eigenfunctions}). These very rapid oscillations are most evident in the evolution of $\rho$ and $u_x$ but are also seen in other components $u_y$, $b_y$, $b_x$ and $b_z$, although the latter primarily oscillate with a lower frequency due to the magnetic mode and newly generated IMWs (Fig. \ref{fig:evolution}). The energy needed for the SDW and IMW excitation is mainly extracted from the shear flow by the magnetic mode and transferred to SDWs and IMWs. 

Then, $k_x(t)$ increases and moves in the next  adiabatic region $k_x(t)/k_y\gg 1$. The magnetic mode leaves the MRI-active region and the linear dynamics of SDWs, IMWs  and the magnetic
mode as well as the counter-propagating components for each of these modes become decoupled again, since all the elements of ${\bf M}$ are small again, tending to zero with increasing $|k_x(t)|$. As a result, there is no further energy exchange among the modes themselves and between the modes and the flow. The time-scales of the modes
get separated again: $\psi_s^{(\pm)}$ and the associated $\rho$ and $u_x$ rapidly oscillate (Figs. \ref{fig:evolution} and \ref{fig:eigenfunctions}) because of the newly excited SDWs with the
frequency $\omega_s$ linearly increasing with time,  while IMWs and the magnetic mode eigenfunctions, $\psi_{im}^{(\pm)}$ and $\psi_m^{(\pm)}$, continue to oscillate with the respective $\omega_{im}$ and $\omega_m$ frequencies (Fig. \ref{fig:eigenfunctions}), which remain constant in time, $\omega_m\approx \omega_{im}\approx v_Ak_z$. The other velocity and magnetic field components, being mainly related to the magnetic mode and IMWs,  are accordingly dominated by these low-frequency oscillations, although still contain high-frequency contribution arising from SDWs whose amplitude are, however, much smaller than that of the low-frequency parts (Fig. \ref{fig:evolution}), except in $b_z$ where the contribution of rapid oscillations is larger (see equation \ref{eq:bzkk}). Note that in the incompressible case, at large $k_x\rightarrow \infty$, the longitudinal component $u_x$ associated with the compressible SDWs are zero, since these waves are absent in that case, while the perpendicular components $u_y,u_z$ and $b_y,b_z$ are nonzero and are associated with the incompressible magnetic and IMWs. This is the reason why the high-frequency oscillations of SDWs when they emerge in the present compressible case are most noticeable in $\rho$ and $u_x$. Since our main goal is to study the generation of compressive SDWs by the incompressive MRI-unstable magnetic mode, below we further characterize this process as a function of azimuthal and vertical wavenumebrs.

Finally, we point out that the generation of SDWs by the magnetic mode is mathematically a consequence of the well-known Stokes phenomenon, whereby different asymptotic WKB solutions of a differential equation emerge (i.e., change from being sub-dominant to dominant) as the complex (temporal) variable crosses anti-Stokes lines in the complex plane \citep[][see also HP09a for the case of vortex-SDW coupling]{Heading1962, Berry1972}. Physically, as mentioned above, it is caused by the nonmodal mechanism due to shear, which, changing $k_x$ with time, gives rise to non-adiabatic regions with respect to the real variable (which is $t$ or $k_x=qk_yt$ here), where time-scales of the modes and shear time become comparable to each other.  Here we briefly illustrate this for SDWs, which are of main interest in this paper. 

\begin{figure}
\centering\includegraphics[width=\columnwidth]{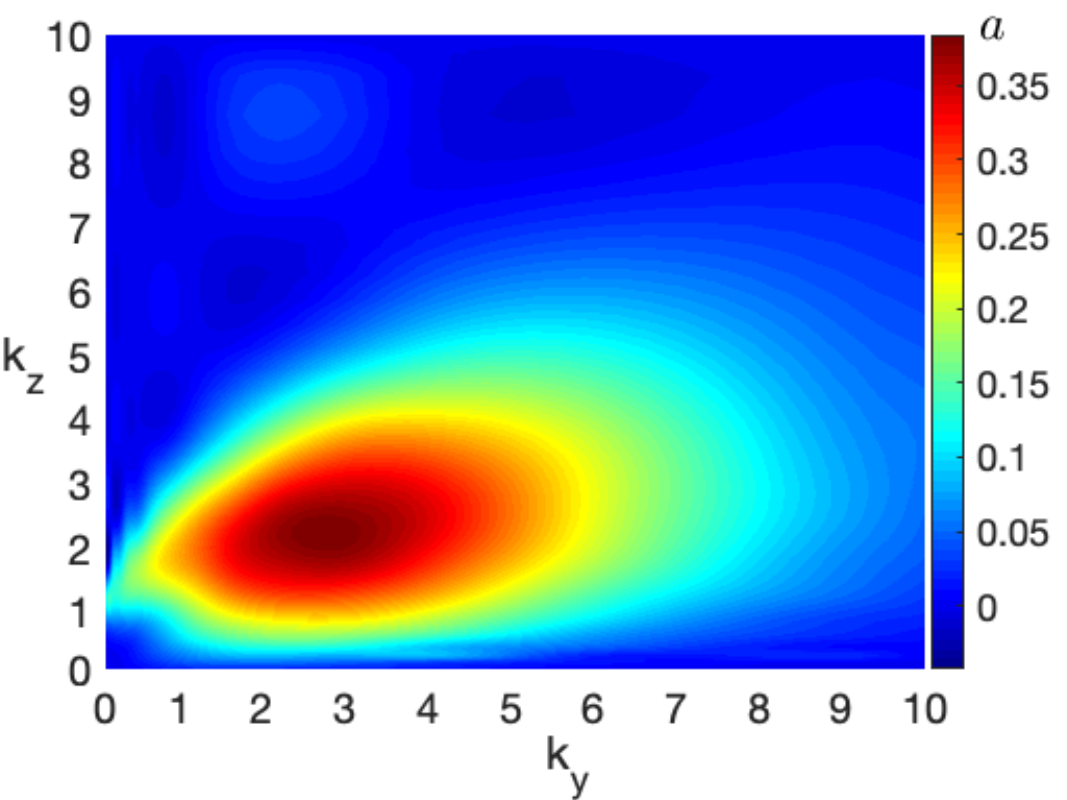}
\caption{The amplitude $a$ of generated SDWs for the density as a function of $k_y$ and $k_z$. It achieves the maximum $a_{m}=0.38$ and, hence the wave generation is most efficient, at $(k_{ym},k_{zm})=(2.8, 2.2)$ and decreases at smaller and larger $k_y$ and $k_z$.}\label{fig:SDWamplitude}
\end{figure}

Let us consider the governing equations (\ref{eq:modal_eqs_time_sp}) and (\ref{eq:modal_eqs_time_sm}) for the eigenfunctions $\psi_s^{(\pm)}$ of SDWs from the full set of coupled modal equations (\ref{eq:modal_eqs_time}). For simplicity, retaining only the dominant eigenfunction $\psi_{m}^{(-)}$ of the initially imposed magnetic mode on the right hand side of these equations, we get 
\begin{equation}
\frac{d\psi^{(+)}_s}{dt}-({\rm i}\omega_s+M_{55})\psi^{(+)}_s=M_{52}\psi^{(-)}_m, \label{eq:psi_sp_short}
\end{equation}
\begin{equation}
\frac{d\psi^{(-)}_s}{dt}+({\rm i}\omega_s-M_{66})\psi^{(-)}_s=M_{62}\psi^{(-)}_m. \label{eq:psi_sm_short}
\end{equation}
In the initial adiabatic region at $t=-\infty$, there are no SDWs and the solutions of equations (\ref{eq:psi_sp_short}) and (\ref{eq:psi_sm_short}) are given by  their particular solutions proportional to $\psi_m^{(-)}$, which accordingly oscillate with $\omega_m$, 
\begin{equation}
\psi_s^{(+)}=-\frac{M_{52}}{{\rm i}(\omega_s+\omega_m)+M_{55}}\psi_m^{(-)},
\end{equation}
\begin{equation}
\psi_s^{(-)}=\frac{M_{62}}{{\rm i}(\omega_s-\omega_m)-M_{66}}\psi_m^{(-)}
\end{equation}
and represent induced oscillations in $\psi_s^{(\pm)}$ by the magnetic mode. Since $M_{52}$ and $M_{62}$ decay at $|t|\rightarrow \infty$ (Fig. \ref{fig:coupling_coeff}), both these solutions also decay with time. As $k_x$ crosses the non-adiabatic region around $t=0$, WKB regime breaks down, so this asymptotic solution is no longer valid and SDWs are excited due to the Stokes phenomenon. As a result, in the next adiabatic region at $t=\infty$, the solution is equal to the  superposition of the excited SDWs oscillating with the high frequency $\omega_s$, which represent homogeneous solution, and the former low-frequency particular solution originating from the magnetic mode (other contributions from IMWs are omitted here for simplicity),
\begin{equation}
\psi_s^{(+)}=a_s^{(+)}e^{\int_0^t[{\rm i}\omega_s(t')+M_{55}(t')]dt'}-\frac{M_{52}}{{\rm i}(\omega_s+\omega_m)+M_{55}}\psi_m^{(-)}, \label{eq:solution_final_psi_sp}
\end{equation}
\begin{equation}
\psi_s^{(-)}=a_s^{(-)}e^{\int_0^t[{-\rm i}\omega_s(t')+M_{66}(t')]dt'}+\frac{M_{62}}{{\rm i}(\omega_s-\omega_m)-M_{66}}\psi_m^{(-)}, \label{eq:solution_final_psi_sm}
\end{equation}
where $a_s^{(+)}$ and $a_s^{(-)}$ are the corresponding constant amplitudes of the excited counter-propagating SDWs. It follows from the form of the coupling matrix $\bf M$ (equation \ref{eq:M}) that at $|t|\rightarrow \infty$, the real parts ${\rm Re}(M_{55})={\rm Re}(M_{66})=-4.5\omega_s^{-1}(d\omega_s/dt)=-4.5d~{\rm ln}\omega_s/dt$ and hence the SDW parts in solutions (\ref{eq:solution_final_psi_sp}) and (\ref{eq:solution_final_psi_sm}), after time integration of $M_{55}$ and $M_{66}$ in the exponent, scale as $\omega_s^{-4.5}$, which differs from the standard WKB-type dependence $\omega_s^{-1/2}$ (HP09a). This rapid decrease of the SDW parts at large times in $\psi_s^{(\pm)}$ is also seen in Fig. \ref{fig:eigenfunctions}. In the standard situation, the WKB approach is applied to a second order differential equation and asymptotic solutions at $t=-\infty$ and $t=\infty$ are matched to find the amplitudes of generated waves. In particular, for the vortex-wave coupling in the simplest 2D case ($k_z=0$), HP09a analytically found the amplitude of generated SDWs by the aperiodic vortical mode by matching WKB solutions across the anti-Stokes lines in the complex variable plane. However, in the present case analytically finding these amplitudes is complicated because of the complex form of a full set of modal equations (\ref{eq:modal_eqs_time}) involving a larger number of interacting modes -- the magnetic mode, IMWs and SDWs each with two counter-propagating components. For this reason, we adopt a different approach and quantify the SDW amplitudes -- and hence their generation efficiency -- using the density perturbation, in which, as noted above, these waves are most clearly manifested, rather than the asymptotic forms (\ref{eq:solution_final_psi_sp}) and (\ref{eq:solution_final_psi_sm}) of the SDW eigenfunctions.

\subsection{Generation of SDWs by MRI for various $k_y$ and $k_z$}

Figure \ref{fig:density_diffkykz} shows the evolution of the density for the initially imposed magnetic mode as in Fig. \ref{fig:evolution}, but for different $k_y$ and $k_z$. For large $k_z=10$  (i.e., when the
vertical wavelengths of the modes are much smaller than the disc scale height), there are no rapid oscillations, implying that SDW generation does not occur even at higher $k_y=10$; there are only low-frequency oscillations induced by the original magnetic mode and excited IMWs. The absence of SDWs in this case of large $k_z\gg k_y$ is because of the fact that the time-scales of SDWs and the magnetic  mode remain well separated during an entire course of the evolution and the WKB regime is uniformly valid at all times. A similar behavior is also observed at $k_y=0.1$ and $k_z=1$. By contrast, an efficient SDW generation takes place at smaller $k_z=0.1, 1$ and larger $k_y=1, 10$, or more generally $k_z\lesssim k_y$, where the favourable conditions for the mode coupling arise -- existence of the non-adiabatic region at $|k_x(t)|\lesssim k_y$ where the time-scales of the SDW, IMWs, magnetic mode and shear time $1/q$ are comparable. This leads to an efficient generation of SDWs by the magnetic mode evidenced by the presence of high-frequency oscillation with increasing amplitude in $\rho$. A moderate SDW generation is also observed at small $k_y=k_z=0.1$.

We can quantify the efficiency of the SDW generation by calculating wave amplitudes as a function of $k_y$ and $k_z$. Taking into account the asymptotic expressions for SDW parts in $\psi_s^{(\pm)}$, which scale as $\omega_s^{-4.5}$, from equations (\ref{eq:solution_final_psi_sp}) and (\ref{eq:solution_final_psi_sm}) and the form of the transformation matrix $\bf C$ (equation B1) at large times $t\rightarrow \infty$, we obtain the asymptotic form of the high-frequency part of the density associated with SDWs  
\begin{equation}
\rho=\omega_s^{1/2}(t)\left(a_1e^{{\rm i}\int_0^t\omega_s(t')dt'}  + a_2e^{-{\rm i}\int_0^t\omega_s(t')dt'}\right), \label{eq:density_SDW_part}
\end{equation}
whose magnitude thus increases with time as $\omega_s^{1/2}(t)$, as seen in Fig. \ref{fig:evolution}. Here $a_1$ and $a_2$ are the constant in time complex amplitudes of each counter-propagating SDW branch near the generation point $t=0$. Figure \ref{fig:SDWamplitude} shows the dependence of the total amplitude $a=(|a_1|^2+|a_2|^2)^{1/2}$ of the waves, on $k_y$ and $k_z$. It is seen that when these wavenumbers are large, $|a|$ is small and therefore the SDW generation is inefficient consistent with the behaviour in Fig. \ref{fig:density_diffkykz}. With decreasing $k_y$ and $k_z$, the SDW amplitude increases and attains its maximum value $a_{m}=0.38$ at $(k_{y,m},k_{z,m})=(2.8, 2.2)$, indicating that the most efficient generation of non-axsymmetric SDWs by the magnetic mode takes place at azimuthal and vertical wavelength, $\lambda_{y,m}=2\pi/k_{y,m}=2.24$ and $\lambda_{z,m}=2\pi/k_{z,m}=2.9$, of perturbations comparable to the disc scale height $H$. This is consistent with the results of HP09a who showed that the SDW excitation by the vortical mode in the 2D case ($k_z=0$) is most efficient  at comparable, but somewhat smaller $k_{y,m}=0.78$ (in units of $H^{-1}$), however, in the present case of SDW generation by the magnetic mode, it appears to be more efficient at higher $k_z\gtrsim  1$. 
The values of $k_{y,m}, k_{z,m}$ lie in the region $1\lesssim k_y,k_z \lesssim 3$ where non-axisymmetric magnetic mode undergoes appreciable nonmodal MRI growth, as shown in \citet[][see their Figs. 1 and 2]{Gogichaishvili_etal2018}. This indicates that SDWs are excited most efficiently at wavenumbers for which MRI is still significant.

\section{Summary and discussion}

In this paper, we have revealed and investigated a linear process of generation of spiral density waves by MRI-unstable non-axisymmetric magnetic perturbations in compressible Keplerian discs with a nonzero net vertical background magnetic field in the shearing box approach. This wave excitation process is a consequence of a generic linear mode coupling phenomenon associated with the nonmodal dynamics of perturbations arising from the non-self-adjoint nature of shear flows, including  Keplerian discs.

As is usually done in the shearing box, we have decomposed perturbations into spatial Fourier harmonics -- shearing plane waves with a time-dependent radial wavenumber $k_x$ -- and analyzed their evolution and dynamics by numerically solving  the linearized MHD equations for different azimuthal, $k_y$, and vertical, $k_z$, wavenumbers. 
We have derived the general dispersion relation in the WKB, or adiabatic regime, $k_y\ll k_z$, in order to classify perturbation modes in the considered flow system. These modes are: SDWs driven mainly by compressibility and two incompressible modes -- IMWs driven by rotation and magnetic tension force and the magnetic mode driven by magnetic tension. The latter two modes become standard Alfv\'en waves at large $k_x$. The magnetic mode plays a key dynamical role, since it is the mode that exhibits nonmodal MRI growth at smaller $k_x$. Using the canonical formulation, we have rewritten the main linearized MHD equations for the eigenfunctions of these modes first in the WKB regime when the equations diagonalize and the mode eigenfunctions decouple. Then, we have generalized these modal equations to the more important non-adiabatic regime, when they become coupled for non-aixsymmetric ($k_y\neq 0$) eigenfunctions due to shear, giving rise to dynamical couplings among these three modes. Then, we have focused on the coupling of non-axisymmetric SDWs and the magnetic mode. To this end, we  initially imposed the magnetic mode of a tightly leading orientation ($k_x/k_y<0,~|k_x|\gg k_y,k_z)$ in the adiabatic regime and traced the subsequent evolution of density, velocity and magnetic field perturbations as well as mode eigenfunctions till the magnetic mode eventually becomes tightly trailing ($k_x/k_y>0, k_x\gg k_y$) in the next adiabatic region. The magnetic mode, evolving in time, enters the non-adiabatic coupling interval ($|k_x|\lesssim k_y$), where it undergoes transient, or nonmodal MRI  growth and, at crossing the point $k_x=0$, abruptly generates SDWs and IMWs. The generated SDWs are evidenced by the appearance of high-frequency oscillations notably in the density and radial velocity perturbations, while the other components of velocity and magnetic field, being related to the magnetic mode and IMWs, oscillate with lower, of the order of the Alfv\'en frequency. The mode coupling in the non-adiabatic region occurs because the coupling terms between different mode branches become important due to shear and the characteristic time-scales of these modes come close to each other, enabling an efficient mode interaction in this interval.

We have also characterized the SDW generation as a function of $k_y$ and $k_z$ and found that it is most efficient when these wavenumbers $k_y,k_z\sim H^{-1}$, that is,  when the azimuthal and vertical wavelengths of perturbations are comparable to the disc scale height. At these wavenumbers nonmodal MRI growth for non-axisymmetric magnetic mode is still appreciable, implying that the regimes of MRI and SDW generation overlap. 

Because the strongest MRI–SDW coupling occurs at wavelengths comparable to the disc scale height, vertical stratification may therefore play a dynamical role. For the sake of mathematical simplicity, we have ignored vertical gravity and the resulting stratification of density and pressure, assuming that they are uniform in the vertical $z$-direction. This simplification allowed us to isolate and focus on the essential mechanism of the shear-induced generation of SDWs by the MRI-unstable magnetic mode. As shown in \cite{Mamatsashvili_etal2013}, stratification primarily modifies IMW mode, transforming it into inertia-gravity wave, while only weakly affects the  magnetic mode, leaving its growth rate due to MRI essentially unchanged. Thus, although stratification may quantitatively modify the mode frequencies and eigenfunctions, it is not expected to alter the fundamental dynamics of the SDW generation process studied here. This process, being a consequence of the shear-induced linear mode coupling phenomenon, remains robust as it originates from the non-self-adjoint nature of the disc’s differential rotation. The SDW generation occurs mainly in the horizontal $(k_x,k_y)$-plane, where the effect of shear is concentrated, during a short time interval when $k_x(t)\approx 0$, although it still depends on $k_z$. By contrast, the effect of stratification is concentrated mainly in $z$. This timescale of the wave generation process is much shorter than the buoyancy oscillation period associated with stratification.

The present analysis can be generalized to a vertically stratified case within the same shearing box approach, although it would be mathematically more involved. Specifically, in the presence of isothermal stratification, which yields a Gaussian density profile in $z$ and is often adopted in MRI-turbulence studies \citep[e.g.,][]{Davis_etal10, Guan_Gammie2011, Bodo_etal2014, Ryan_etal2017, Held_etal2024}, a linear analysis of MRI can be conducted by expanding the variables in $z$ on the basis of Whittaker cardinal  functions instead of Fourier modes \citep{Latter_etal2010}, while in the horizontal $x$- and $y$-directions the same decomposition in shearing waves can be applied. The order of the Whittaker functions, which is an integer number, determines the vertical scale of the perturbations and therefore plays a role analogous to that of the vertical wavenumber $k_z$ in the unstratified case. It replaces $k_z$ in the governing linear equations and the corresponding dispersion relation, as shown in \cite{Latter_etal2010}. Thus, the main modification introduced by vertical stratification is that shearing waves of perturbation modes are characterized by discrete vertical mode numbers, while retaining a time-dependent $k_x(t)$ and constant $k_y$. Therefore, these stratified shearing waves are expected to exhibit qualitatively similar dynamics -- nonmodal MRI growth and mode coupling -- to those in the unstratified case studied in this paper. In particular, the dependence of the amplitude of the generated SDWs on $k_y$ and vertical mode number in the stratified case will be qualitatively similar to that in Fig. \ref{fig:SDWamplitude}, but, of course, with some quantitative differences.

Finally, a connection of the present results with studies of MRI-turbulence in compressible discs is in order. SDWs have often been observed in simulations of MRI-turbulence with a nonzero net magnetic flux, where they typically appear as large-scale structures, comparable to the box size, in the density distribution and also contribute to Reynolds stress \citep{Fromang_Stone2009, Nelson_Gressel2010,Blaes_etal2011,Flock_etal2011, Gressel_etal2012,Bai_Stone2013,Zhu_etal2013,Ross_Latter2018,Flock_etal2017,Sun_Bai2021}. The production of SDWs and associated density fluctuations by MRI-turbulence have been attributed in those studies to the linear vortex-wave coupling mechanism of HP09a. As noted above, this mechanism is essentially hydrodynamic in nature, relies on the conservation of potential vorticity, and is two-dimensional, i.e. uniform in the vertical direction, as was also confirmed by the same authors in their subsequent nonlinear simulations of zero net flux MRI-turbulence \citep{Heinemann_Papaloizou2009b}. However, in the 3D MHD case with nonzero net flux, the situation is fundamentally different in that potential vorticity is no longer conserved and the MRI-unstable magnetic mode dominates the dynamics, complicating the identification of the pure vortical mode in the turbulent state. As a result, SDWs excited by MRI-turbulence are not always uniform, but rather exhibit some large-scale variation in the vertical direction. The generation of SDWs by the magnetic mode demonstrated in this paper can be regarded as another main process that can produce compressible motions via MRI-driven mostly incompressible velocity perturbations in nonzero net flux MRI-turbulence. For this reason, it is expected that SDW amplitudes and resulting density fluctuations will be stronger than those produced by vortices, although to verify this further nonlinear study of compressible nonzero net flux MRI is required. This may be particularly important for disc heating, since SDWs generated in MRI-turbulence can steepen into shocks, thereby dissipating energy \citep{Rafikov2016, Kaul_etal2025}. The stronger the wave  amplitude at the generation, the more intensive the heating is due to shock dissipation. In addition, the azimuthal and vertical wavelength of the highest wave generation efficiency identified in this study will be important in guiding future nonzero net flux MRI-turbulence simulations to optimally capture SDWs generation process.

\section{acknowledgments}

This work was supported by the Deutsche Forschungsgemeinschaft (DFG) (Grant No. MA10950/1-1) and Shota Rustaveli National Science
Foundation of Georgia (SRNSFG) (Grant No. GHZ-24-047). We thank the anonymous Reviewer for useful comments, which improved the presentation of the paper.

\section{Data availability}
The data underlying this article will be shared on a reasonable request to the corresponding author.

\bibliographystyle{mnras}
\bibliography{biblio}

\appendix
\onecolumn

\section{Expression for the evolution matrix $A$}

The evolution matrix $\bf A$ in equations (\ref{eq:matrix_eqs}) and (\ref{eq:matrix_eqs_time}) has the following form

\begin{equation}
\mathbf{A}=
\begin{pmatrix}
0 & 1 & 0 & 0 & 0 & 0\\ \\
-k_z^2 & 0 & k_xk_z & 0 & k_yk_z & 0\\ \\
0 & 0 & 0 & 1 & 0 & 0\\ \\
k_xk_z & 0 & 2q-\frac{2k_z^2}{\beta}-\left(1+\frac{2}{\beta}\right)k_x^2& 0 & -\left(1+\frac{2}{\beta}\right)k_xk_y  & 2 \\ \\
0 & 0 & 0 & 0 & 0 & 1\\ \\
k_yk_z & 0 & -\left(1+\frac{2}{\beta}\right)k_xk_y& -2 & -\frac{2k_z^2}{\beta}-\left(1+\frac{2}{\beta}\right)k_y^2  & 0
\end{pmatrix},
\end{equation}
which is readily obtained by substituting the new variables 
$(h_1, h_2, h_3, h_4, h_5, h_6)$ introduced in Section 2.2 in equations (\ref{eq:bxkk})-(\ref{eq:bzkk}).

\section{Expression for the transformation matrix $C$}

The eigenvectors of the evolution matrix ${\bf A}$ are obtained by substituting its eigenvalues $\pm {\rm i}\omega_m$, $\pm {\rm i}\omega_{im}$, $\pm{\rm i}\omega_s$, as given by dispersion relation (\ref{eq:dispersion}), in equation (\ref{eq:modal_eqs}) and, for a chosen normalization, have the form $[f_1(\omega),~{\rm i}\omega f_1(\omega),~f_2(\omega),~{\rm i}\omega f_2(\omega),~ f_3(\omega),~{\rm i}\omega f_3(\omega)]^T$, where $\omega=\pm\omega_m,~\pm\omega_{im},~\pm \omega_s$ and the functions $f_1, f_2$ and $f_3$ are
\[
f_1(\omega)=\frac{k_z[k_xf_2(\omega)+k_yf_3(\omega)]}{k_z^2-\omega^2},
\]
\[
f_2(\omega)=\omega^4-(k_y^2+k_z^2)\left[\omega^2\left(1+\frac{2}{\beta}\right)-\frac{2k_z^2}{\beta}\right], 
\]
\[
f_3(\omega)= 2{\rm i}\omega(\omega^2-k_z^2)+  k_xk_y\frac{\omega^4-f_2(\omega)}{k_y^2+k_z^2}.
\]
The columns of the transformation matrix $\bf C$, as defined in the Section 2.2, are constructed from the eigenvectors of $\bf A$ and can be written in the following order 
\begin{equation}
\mathbf{C}=
\begin{pmatrix}
f_1(\omega_m) & f_1(-\omega_m) & f_1(\omega_{im}) & f_1(-\omega_{im}) & f_1(\omega_s) & f_1(-\omega_s)\\ \\
{\rm i}\omega_mf_1(\omega_m) & -{\rm i}\omega_mf_1(-\omega_m) & {\rm i}\omega_{im}f_1(\omega_{im}) & -{\rm i}\omega_{im}f_1(-\omega_{im}) & {\rm i}\omega_sf_1(\omega_s) & -{\rm i}\omega_sf_1(-\omega_s)\\ \\
f_2(\omega_m) & f_2(-\omega_m) & f_2(\omega_{im}) & f_2(-\omega_{im}) & f_2(\omega_s) & f_2(-\omega_s)\\ \\
{\rm i}\omega_mf_2(\omega_m) & -{\rm i}\omega_mf_2(-\omega_m) & {\rm i}\omega_{im}f_2(\omega_{im}) & -{\rm i}\omega_{im}f_2(-\omega_{im}) & {\rm i}\omega_sf_2(\omega_s) & -{\rm i}\omega_sf_2(-\omega_s)\\ \\
f_3(\omega_m) & f_3(-\omega_m) & f_3(\omega_{im}) & f_3(-\omega_{im}) & f_3(\omega_s) & f_3(-\omega_s)\\ \\
{\rm i}\omega_mf_3(\omega_m) & -{\rm i}\omega_mf_1(-\omega_m) & {\rm i}\omega_{im}f_3(\omega_{im}) & -{\rm i}\omega_{im}f_3(-\omega_{im}) & {\rm i}\omega_sf_3(\omega_s) & -{\rm i}\omega_sf_3(-\omega_s). 
\end{pmatrix}
\end{equation}

It can be verified that with this form of $\bf C$, the spectral decomposition $\bf A=\bf C \Lambda \bf C^{-1}$, holds, which is used in the main analysis. As before,   $\Lambda$ is a diagonal matrix composed of the eigenvalues of $\bf A$, i.e., ${\rm i}\omega_m,~-{\rm i}\omega_m,~{\rm i}\omega_{im},~-{\rm i}\omega_{im},~{\rm i}\omega_s,~-{\rm i}\omega_s$, which are ordered corresponding to the columns of $\bf C$,

\begin{equation}
\mathbf{\Lambda}=
\begin{pmatrix}
{\rm i}\omega_m & 0 & 0 & 0 & 0 & 0 \\ 
 0 & -{\rm i}\omega_m & 0 & 0 & 0 & 0 \\ 
 0 & 0 & {\rm i}\omega_{im} & 0 & 0 & 0 \\ 
 0 & 0 & 0 & -{\rm i}\omega_{im} & 0 & 0 \\ 
 0 & 0 & 0 & 0 & {\rm i}\omega_s & 0 \\ 
 0 & 0 & 0 & 0 & 0 & -{\rm i}\omega_s \\ 
\end{pmatrix}
\end{equation}

\end{document}